# Texture Formation in Polycrystalline Thin Films of All-Inorganic Lead Halide Perovskite

*Julian A. Steele,\* Eduardo Solano, Handong Jin, Vittal Prakasam, Tom Braeckevelt, Haifeng Yuan, Zhenni Lin, René de Kloe, Qiong Wang, Sven M. J. Rogge, Veronique Van Speybroeck, Dmitry Chernyshov, Johan Hofkens, and Maarten B. J. Roeffaers*

Dr. J. A. Steele, Dr. V. Prakasam, Prof. M. B. J. Roeffaers
cMACS, Department of Microbial and Molecular Systems, KU Leuven, 3001 Leuven, Belgium
Email: julian.steele@kuleuven.be

Dr. E. Solano
NCD-SWEET beamline, ALBA synchrotron light source, 08290, Cerdanyola del Vallès, Barcelona, Spain

H. Jin, T. Braeckevelt, Dr. H. Yuan, Prof. J. Hofkens
Department of Chemistry, KU Leuven, 3001 Heverlee, Belgium

T. Braeckevelt, Dr. S. M. J. Rogge, Prof. V. Van Speybroeck,
Center for Molecular Modeling (CMM), Ghent University, Technologiepark 46, 9052 Zwijnaarde, Belgium

Z. Lin
Materials Sciences Division, Lawrence Berkeley National Laboratory, Berkeley, CA 94720, United States

Z. Lin
Department of Materials Science and Engineering, University of California, Berkeley, CA 94720, United states

Dr. R. de Kloe
EDAX, Ametek BV, Ringbaan Noord 103, 5046 AA Tilburg, The Netherlands

Dr. Q. Wang
Young Investigator Group Active Materials and Interfaces for Stable Perovskite Solar Cells, Helmholtz-Zentrum Berlin für Materialien und Energie, 12489 Berlin, Germany

Dr. D. Chernyshov
Swiss-Norwegian Beamlines at the European Synchrotron Radiation Facility, 71 Avenue des Martyrs, F-38043 Grenoble, France

Prof. J. Hofkens
Max Plank Institute for Polymer Research, Mainz, D-55128, Germany

**Abstract:** Controlling grain orientations within polycrystalline all-inorganic halide perovskite solar cells can help increase conversion efficiencies toward their thermodynamic limits, however the forces governing texture formation are ambiguous. Using synchrotron X-ray diffraction, we report meso-structure formation within polycrystalline CsPbI$_{2.85}$Br$_{0.15}$ powders as they cool from a high-temperature cubic perovskite (α-phase). Tetragonal distortions (β-phase) trigger preferential crystallographic alignment within polycrystalline ensembles, a feature we suggest is coordinated across multiple neighboring grains via interfacial forces that select for certain lattice distortions over others. External anisotropy is then imposed on polycrystalline thin films of orthorhombic (γ-phase) CsPbI$_{3-x}$Br$_x$ perovskite via substrate clamping, revealing two fundamental uniaxial texture formations; (i) I-rich films possess orthorhombic-like texture (<100> out-of-plane; <010> and <001> in-plane), while (ii) Br-rich films form tetragonal-like texture (<110> out-of-plane; <1-10> and <001> in-plane). In contrast to relatively uninfluential factors like the choice of substrate, film thickness and annealing temperature, Br incorporation modifies the γ-CsPbI$_{3-x}$Br$_x$ crystal structure by reducing the orthorhombic lattice distortion (making it more tetragonal-like) and governs the formation of the different, energetically favored textures within polycrystalline thin films.

Lead halide perovskite semiconductors have become a prominent part of solar cell research in recent years due to their unique capacity to combine high-quality optoelectronic performance[1] (e.g. broadband, intense optical absorption and long carrier diffusion lengths) with the ability to cheaply and easily solution-process them into polycrystalline thin films.[2] At present, single-junction solar cells with conversion efficiencies over 25% have been realized, surpassing 29% in tandem cells[3,4] Improving the performance of halide perovskite devices has conventionally taken the form of passivating traps,[3,5] controlling grain size[6] and morphology,[7] and engineering transport layers with optimal interfacing and band alignment.[8] Recently, regulating the direction and distribution of grain orientations within thin films, i.e. so-called crystal texture, has been recognized as a key parameter in need of further optimization.[9–13] This is because the network of fine grains making up solution-processed halide perovskite thin films suffer from spatial heterogeneity in their (opto)electronic properties.[14–16] Beyond identifying processing parameters which can influence grain orientations and their distributions (e.g.

material composition,[17–19] annealing temperature[20,21] and annealing time[22]), the underlying origins of texture formation in halide perovskite thin films are still unclear.

The coordinated alignment of structural domains across multiple halide perovskite crystals requires two central ingredients: (i) intrinsic domain formation across multiple grains, either through widespread nucleation or rapid spreading, and (ii) introduction of an external anisotropic field which can preferentially select for the formation of certain domains over others.[23] Due to their improved thermal and chemical stability,[24] all-inorganic cesium lead halide perovskites are envisioned to champion long-term, stable devices. However, $CsPbI_3$-based perovskites form a stable yellow non-perovskite structure (δ-phase) at room temperature (RT), requiring high-temperature processing (>320°C) to access their black perovskite phases[25] (cubic, α; tetragonal, β; orthorhombic, γ) before being stabilized for applications at RT.[26] Starting from the archetypical high-temperature cubic phase in Figure 1A, octahedral distortions reduce the lattice symmetry and form structural domains[27] upon cooling, being driven by thermal contraction and the development of spontaneous strains within the crystal.[28] High temperature processing of halide perovskite is typically paralleled by large changes in the final thin film (micro)structure and morphology. For instance, annealing induces grain coarsening (Figure 1B) which is often tailored toward evolving smaller nanocrystals into larger micron-scale grains, to reduce the parasitic influence of grain boundaries.[7]

First-order high-temperature phase transitions within halide perovskites[28] unavoidably imply phase intergrowths and co-existence. Within this context, it is their inherent polymorphic nature[28] and relatively soft crystal structure (tendency to form heterojunctions) which will govern complex meso-structure formations, like the thin film texture illustrated in Figure 1C. Thus, a relatively simple perovskite crystal structure may exhibit immense structural complexity at larger length scales, influencing important features – like micrometer carrier diffusion lengths[29,30] – at the thin film-level.

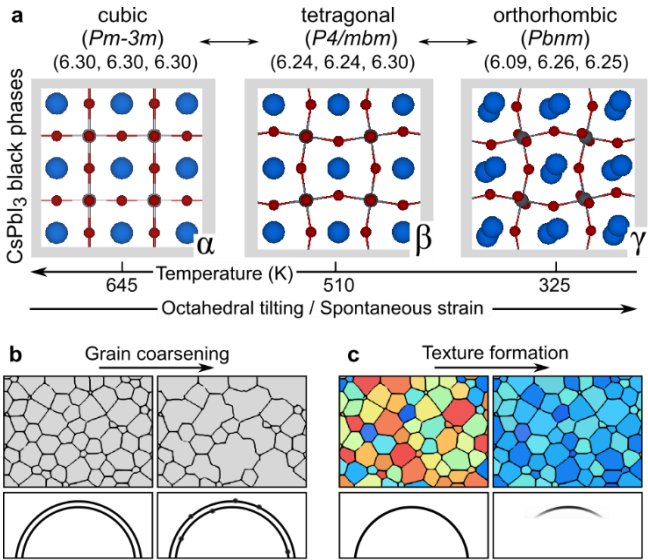

**Figure 1.** (a) Crystal structure of CsPbI$_3$ perovskite (Cs: blue, Pb: black, I: red) between metastable RT orthorhombic structure and a high-temperature cubic, as derived from Marronnier et al.[31] The numbers in brackets indicate the normalized lattice parameters (*a,b,c*) taken from Ref [31], in units of Ångströms. (b) and (c) show the different thermal-driven grain growth processes and grain texture formation in thin films, respectively. Here the upper images show top-down illustrations of thermal-induced changes to thin-film grain microstructure. The black and white diagrams along the bottom schematically show how these two different structural evolutions are reflected in the XRD data collected with a large-area X-ray detector. Colors in (c) ranging from blue to red indicate the relative degree of grain misorientation qualitatively.

In this communication, we examine how changes in the crystal symmetry of free-standing polycrystalline all-inorganic halide perovskites drive common grain orientations and anisotropy during annealing and subsequent cooling, manifesting as texture within polycrystalline thin films. Using synchrotron X-ray diffraction (SXRD) we decouple the development of crystallinity (i.e., the enlargement of crystal grains) and domain ordering (i.e., the collective alignment of octahedral tilts within separate grains) in ensembles of CsPbI$_{2.85}$I$_{0.15}$ micro-crystalline powders during thermal treatment, which represents two key processes: (i) heating randomly orientated as-grown grains causes them to sinter and coarsen before transitioning to the black perovskite phase, where (ii) cooling of the α-phase perovskite polycrystalline assembly introduces common distortion directions within the individual crystals, triggered by the β-phase transition. For thermally treated γ-CsPbI$_{3-x}$Br$_x$ thin films (formed by quenching to RT), a thermal contraction mismatch with the underlying transparent substrate (i.e. glass) biaxially strains the material[32], promoting two distinct uniaxial texture formations, depending on the Br content. We examine thin film texture using synchrotron grazing incident wide angle X-ray scattering

(GIWAXS) and connect the different texture expressions to a Br-induced suppression of orthorhombic distortions in γ-CsPbI$_{3-x}$Br$_x$ perovskite, a feature which is shown to be dominant over all other processing parameters, i.e. annealing temperature, choice of device substrate, and film thickness.

Needle-like, micro-crystalline powders of δ-CsPbI$_{2.85}$Br$_{0.15}$ are grown by drop casting (see Supporting Information; Figure S1)[25,33] which preserve bulk-like phase properties during high-temperature restructuring, without exposure to extrinsic influences like hetero-interfaces[25] or surface-driven effects[34] (i.e. the change in surface tension occurring in nanocrystals). Substituting 5% of the I sites with Br expands the temperature range for which the metastable perovskite phase exists[25] and can be studied *in situ* during phase restructuring.

Meso-structure formed across large grain populations, i.e. with respect to their size, shape, and crystallographic orientation, can arise over lengths of hundreds of nanometers. For SXRD experiments employing large-area 2D detectors, this information is encoded in the shape, width, and azimuthal angular distribution of Debye-Scherrer diffraction rings (Figure 1B,C). SXRD patterns are recorded while heating CsPbI$_{2.85}$Br$_{0.15}$ powders up to 597 K and cooling back to RT, yielding several phase transitions (Figure S2), as displayed in the *T-t* profile in Figure 2A. The δ-CsPbI$_{2.85}$Br$_{0.15}$ powder first transitions into a black phase at 568 K, with the cubic α-phase maintained until cooling where it undergoes a tetragonal distortion, before again transitioning back to a stable δ-phase.

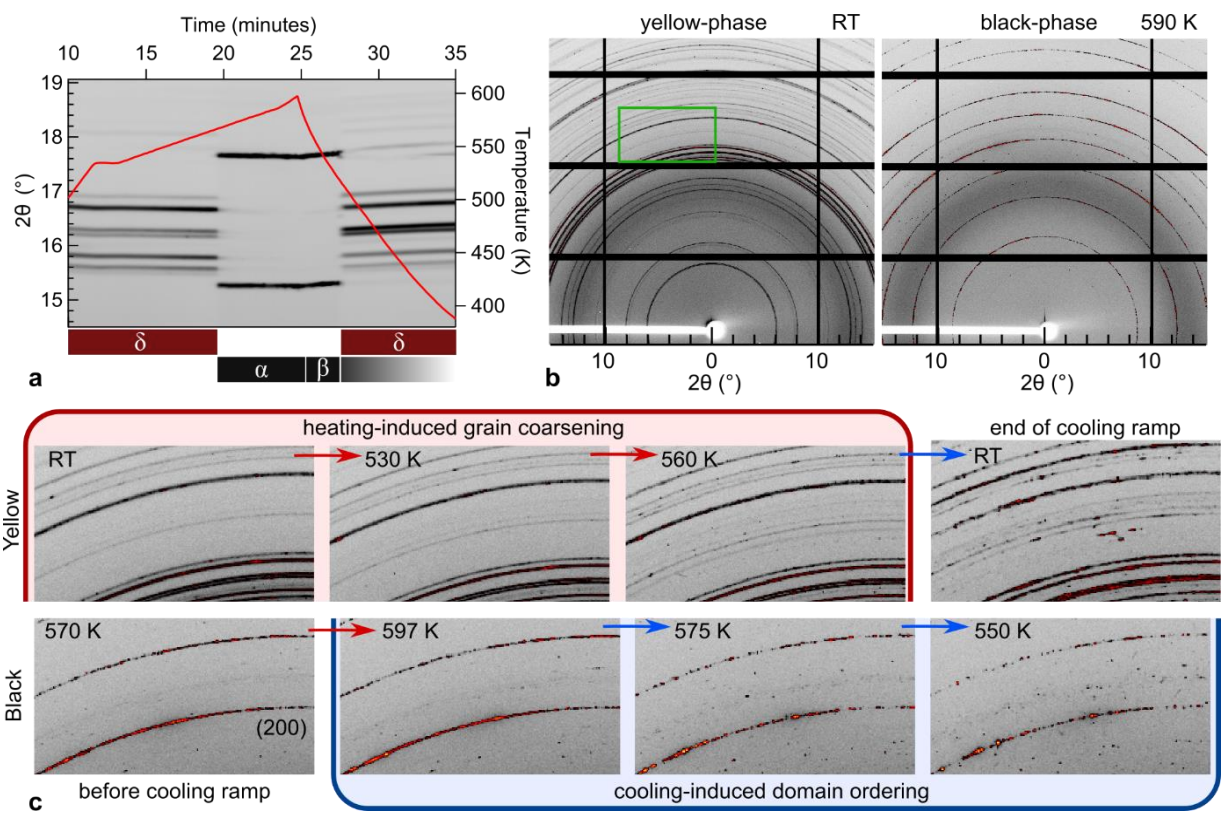

**Figure 2.** (a) In situ SXRD $T$-$t$ profile ($\lambda$ = 0.95774 Å) of CsPbI$_{2.85}$Br$_{0.15}$ through a high-temperature yellow-to-black phase transition (565 K) and cooling ramp. (b) Representative large-area SXRD images recorded from the yellow phase and high-temperature black phase. (c) Enlarged portions of the scatting patterns followed at different stages of the heating (red arrow) and cooling (blue arrow) ramp, corresponding to the green area identified in (B). Note that the images series are separated into yellow (top) and black phase patterns (bottom) at different stages of the same thermal cycle shown in (A).

Figure 2B contains SXRD frames recorded from both the δ-phase and the α-phase before any common grain orientations form (full images in Figure S3), yielding near homogeneous Debye-Scherrer rings due to the inherent disorder of grain orientations at this time. Figure 2C displays an enlargement of the frames to better track the evolution during thermal treatment. Local spottiness develops within the diffraction rings of the yellow phase during heating, indicating grain coarsening (Figure 1B). Thermal-induced changes to micro-crystal morphology is confirmed in Figure S4 which provides scanning electron microscopy (SEM) images of the CsPbI$_{2.85}$Br$_{0.15}$ powders heated to different temperatures, showing widespread sintering of the needle-like microcrystals.

From the yellow-to-black phase transition up to the maximum temperature of 597 K (Figure 2C), no further changes related to ring spottiness and grain coarsening are observed. Cooling, however, induces large changes in the Bragg azimuthal distributions of the SXRD rings; the scattering image recorded at 550 K exhibits a more single-crystal-like pattern, with the formation of highly inhomogeneous and even incomplete scattering rings (i.e. with some pixels in the path of the scattering rings not exposed) and several intense hotspots. Further, regions of high signal intensities are found to bunch and smear, similar to the signals recovered from heavily twinned multiple single crystals. To rule out extensive grain growth as the origin for this rapid change, electron backscatter diffraction (EBSD; Figure S5) measurements confirm that each individual crystal within the partially fused bundles keeps its original microstructure and orientation (i.e. no coherence between grains). This suggests grain growth (Figure 1B) is not the dominant cause of changes within the X-ray scattering ring distribution. Given the relatively large number of individual micro-grains contained in the scattering volume (we estimate as many as ten to twenty thousand grains), such organization is intriguing. We suggest the likely cause is related to the co-alignment of structurally coherent domain blocks at several length scales; from submicron blocks within grains, to large grains forming the polycrystalline aggregates, i.e., microns and tens of microns.

    An azimuthal profile analysis is used to evaluate the distribution of pixel intensities across a portion of the (200) black phase scattering ring during cooling (Figure S6). Cooling from 583 K causes the pixel intensity distribution to shift closer to the detector baseline, as the Debye-Scherrer rings become more intermittent. The degree of inhomogeneity in the intensity distribution across (200) is directly influenced by the size, number and orientation of grains, and is tracked by performing polar integrations over a fragment of the scattering ring (Figure S6). The results in Figure 3A show no development in the azimuthal distribution of the diffracted intensity while heating the black phase (i.e. negligible grain growth), however upon cooling there is a rapid increase in pixel shading (i.e. the fraction of azimuthal bins for which the GIWAXS intensities do not exceed the background). The sudden absence of X-ray signals incident on portions of the (200) Bragg diffraction ring indicates the loss of a good powder average, as one would expect for a randomly oriented polycrystalline material.

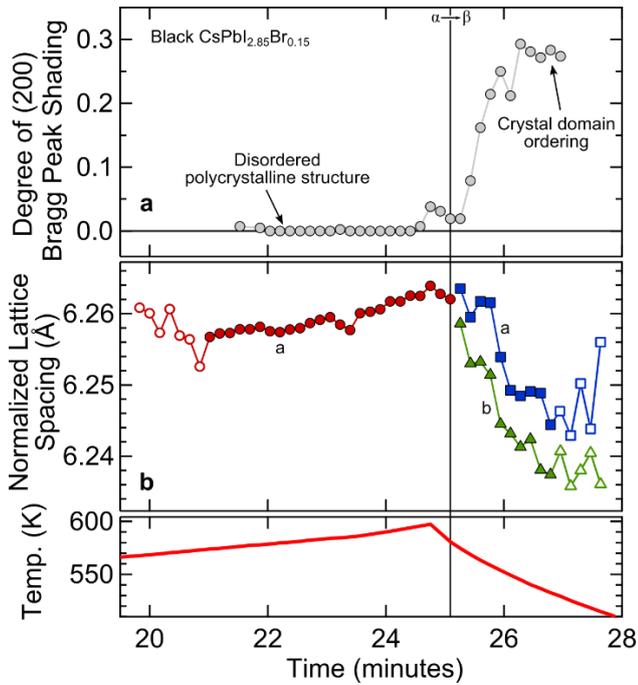

**Figure 3.** (a) Degree of (200) peak shading in black CsPbI$_{2.85}$Br$_{0.15}$ perovskite phase over time, as derived from the Debye ring analysis presented in Figure S6. (b) Normalized lattice parameters of black perovskite phase, with the open shapes representing structural refinements made while in transition with the yellow phase (Figure S2). Normalization of the β-phase lattice is made via rescaling the tetragonal unit cell by (1/√2, , 1/√2, 1). These temporal data both evolve through a corresponding function of temperature, which is shown along the bottom.

An analysis of the normalized lattice spacing evolution reveals that domain ordering correlates with the tetragonal distortion formed in the cooling perovskite unit cell. With β-phase formation appearing to be the trigger, grain orientation will be mediated by inter-granular stimuli within the heavily interfaced polycrystalline network (Figure S4). We suggest that inter-granular tension (the strain exerted by non-coherent interfaces with different thermally expansions[31]), or the highly dynamic nature of the halide perovskite crystals during restructuring,[35] as potential mechanisms for domain selection; however, the preferred directions for grain orientation are not fixed without an external anisotropic field.

We next consider how preferential directions in grain orientations might form under different scenarios. A tetragonal distortion on an isotropic cubic unit cell involves domain formation in one of three equally likely directions (Figure 4A), however the SXRD recorded from free-standing perovskite polycrystalline powders reveal an inherent tendency toward the formation of organized grain orientations during cooling of fused polycrystals (Figure 4B). In the presence of a permeating anisotropic field (i.e. via a strained planar interface), Figure 4C schematically depicts how anisotropic grain orientations are expressed at scale, to form a statistically distributed polycrystalline texture throughout an entire thin film grain population.[25] This we study in detail below.

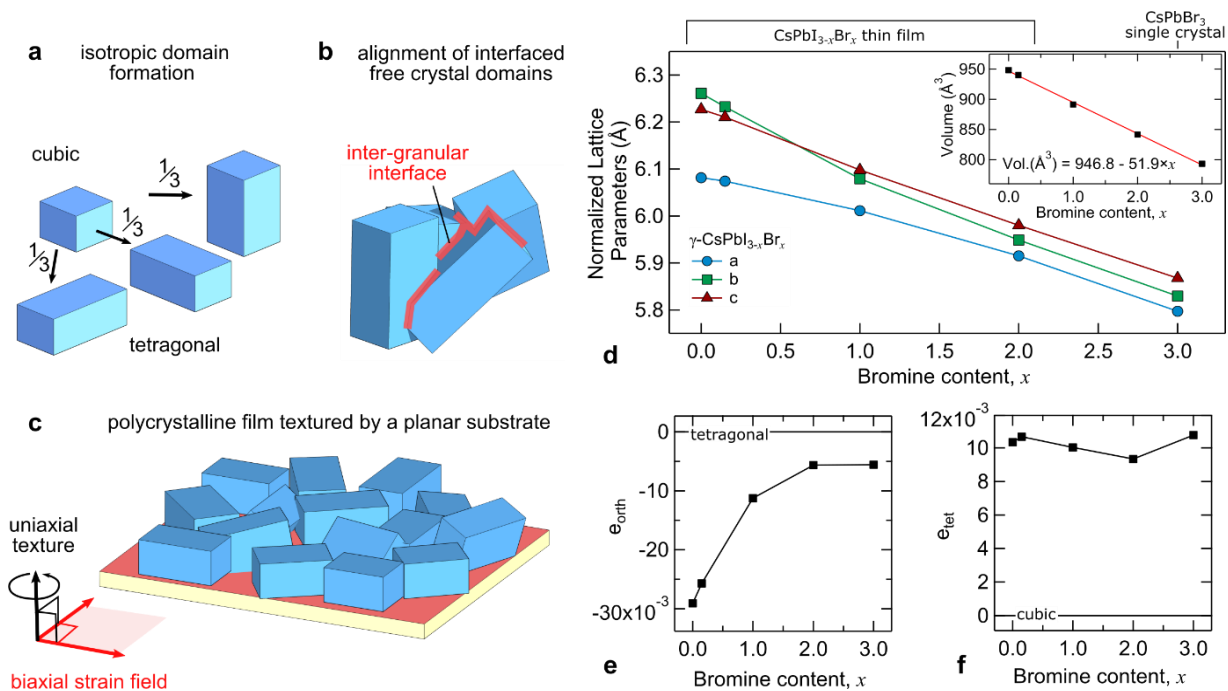

**Figure 4.** Illustrations of crystal orientations within different (poly)crystalline environments; (a) Tetragonal domain formation during the distortion of a cubic unit cell, with equal probability in the three crystallographic directions. (b) Cluster of non-coherent grains sharing a collective preferential alignment of their tetragonal distortion direction. (c) Shared preferential alignment of tetragonal distortions in-plane and the formation of uniaxial texture, due to biaxial substrate strain.[25] For clarity, only domains arising from tetragonal distortions to the cubic unit cell are illustrated. (d) Normalized lattice parameters of γ-CsPbI$_{3-x}$Br$_x$ perovskite as a function of Br content, $x$ (structural refinement shown in Figure S7). Normalization of the γ-phase lattice is made via rescaling the orthorhombic unit cell by (1/√2, 1/√2, 1/2), while the inset shows the unit cell volume evolution which is fit with a straight line. (e) and (f) display the bromine-dependent evolutions of the orthorhombic ($e_{orth}$) and tetragonal ($e_{tet}$) spontaneous strain components, respectively.

With crystal symmetry revealed to drive texture formation, Figure 4D assesses the changes taking place in the normalized lattice of γ-CsPbI$_{3-x}$Br$_x$ perovskite with increasing Br content. A systematic reduction to the crystal volume is imposed by the replacement of I sites with relatively smaller Br anions.[36] The relative changes in the normalized lattice parameters are linked to the redistribution of spontaneous strain measured as deformations to the parent high-symmetry unit cell, induced by the phase transitions.[37] For example, with increasing Br content, the difference between the *a*- and *b*-axes (representing the spontaneous orthorhombic distortion) is significantly reduced. This strongly mirrors the effect of heating a metastable RT γ-CsPbI$_3$ perovskite toward higher temperatures,[31] which increases both the crystal symmetry (becoming more tetragonal-like and then cubic) and the thermodynamic stability.

For phase transitions in which the high symmetry α-phase is reduced to a degenerate γ-phase, the number of distortion components can be expressed in terms of symmetry-adapted strains.[38] The degenerate symmetry-breaking distortions are thus divided into the tetragonal ($e_{tet}$) and orthorhombic ($e_{orth}$) strains, manifesting the β-phase and γ-phase, respectively. These quantities are calculated relative to an undistorted cubic unit cell, $a_0$, which is estimated by taking the cube root of the normalized unit cell volume. It follows that the spontaneous strain components are defined as: $e_1 = (a - a_0)/a_0$, $e_2 = (b - a_0)/a_0$ and $e_3 = (c - a_0)/a_0$, where $a$, $b$ and $c$ are the normalized lattice parameters of the $CsPbI_{3-x}Br_x$ orthorhombic phase. The separate strain components contributing to the lattice distortions are then given by: $e_{orth} = e_1 - e_2$ and $e_{tet} = (2e_3 - e_1 - e_2)/\sqrt{3}$. A factor of $\sqrt{3}$ is included here to ensure that the two strains are on the same scale. Figures 4E and F present the two spontaneous strain components and reveal that $e_{orth}$ is significantly reduced in γ-$CsPbI_{3-x}Br_x$ with rising Br incorporation (~80% for $x \geq 2$), while $e_{tet}$ remains unchanged.

Next we explore texture formation within high-temperature processed polycrystalline γ-$CsPbI_{3-x}Br_x$ thin films (Figure S1) using synchrotron-based GIWAXS. Starting from a randomly-oriented, as-grown δ-$CsPbI_3$ thin film (Figure S8), thermal heating and quenching generates a metastable black phase and introduces biaxial strain[25] due to the large mismatch in thermal expansion of the perovskite layer (~50×10$^{-6}$ K$^{-1}$) and the underlying glass substrate (~4×10$^{-6}$ K$^{-1}$). Figure 5A presents a 2D GIWAXS intensity image of a quenched RT γ-$CsPbI_3$ thin film rendered to correspond with the scattering azimuthal angle (off-center detected image with wider in-plane scattering range is shown in Figure S9). Selected Bragg peak intensity distributions are provided in Figure 5B, showing (002) and (020) reflections reach their maxima in-plane, with (200) pointing normal to the thin film surface. The uniaxial texture confines the (110) peak maximum to point out-of-plane in a bi-modal fashion (two maxima separated by 90°). The azimuthal distribution of (200) – which exhibits no clear peak – is well-described using two adjacent cumulative distribution functions (CDF), to create a step-like probability function centered at $\chi = 0°$: $\varphi_{<100>}(\chi) = A \times [1 - (CDF_+ + CDF_-)]$, where $CDF_\pm = \frac{1}{2}\left[1 + \text{erf}\left(\frac{\pm\chi - \mu}{\sigma\sqrt{2}}\right)\right]$. Here $A$ is a constant, *erf* is the error function, $\mu$ describes the mean deviation from the substrate normal (i.e. $2\mu$ = fwhm), and $\sigma^2$ is the variance at edges, governing the skew rate. Fitting the GIWAXS data recorded from our γ-CsPbI$_3$ thin films with this distribution function, $\varphi_{<100>}(\chi)$, yields $\mu$= 29° and σ=5.1°. We hereon identify the texture found in thin film γ-CsPbI$_3$ as orthorhombic-like, $\varphi_{orth}$; see Figure 5C.

To investigate whether the resultant γ-CsPbI$_3$ texture is a general one, similar analyses were performed on thin films with different thicknesses (Figure S10) and deposited on different inorganic device-ready substrates (Figure S11 and S12). Disregarding relatively rough mesoporous titanium oxide (m-TiO$_2$) substrates (examined below), both the substrate choice (all with thermal expansion coefficients[32,39,40] much lower than the perovskite layer; ranging ~4 – 10×10$^{-6}$ K$^{-1}$) and thin-film thickness (from 140–900 nm) have little influence on the final γ-CsPbI$_3$ thin film texture. This suggests that the texture formation is effectively translated from the strained interface through the whole thin film layer, even when it is comprised of multiple stacked grains, as is the case with our thicker samples. Together, these additional experiments support a crystal symmetry-driven mechanism for the texture formation.

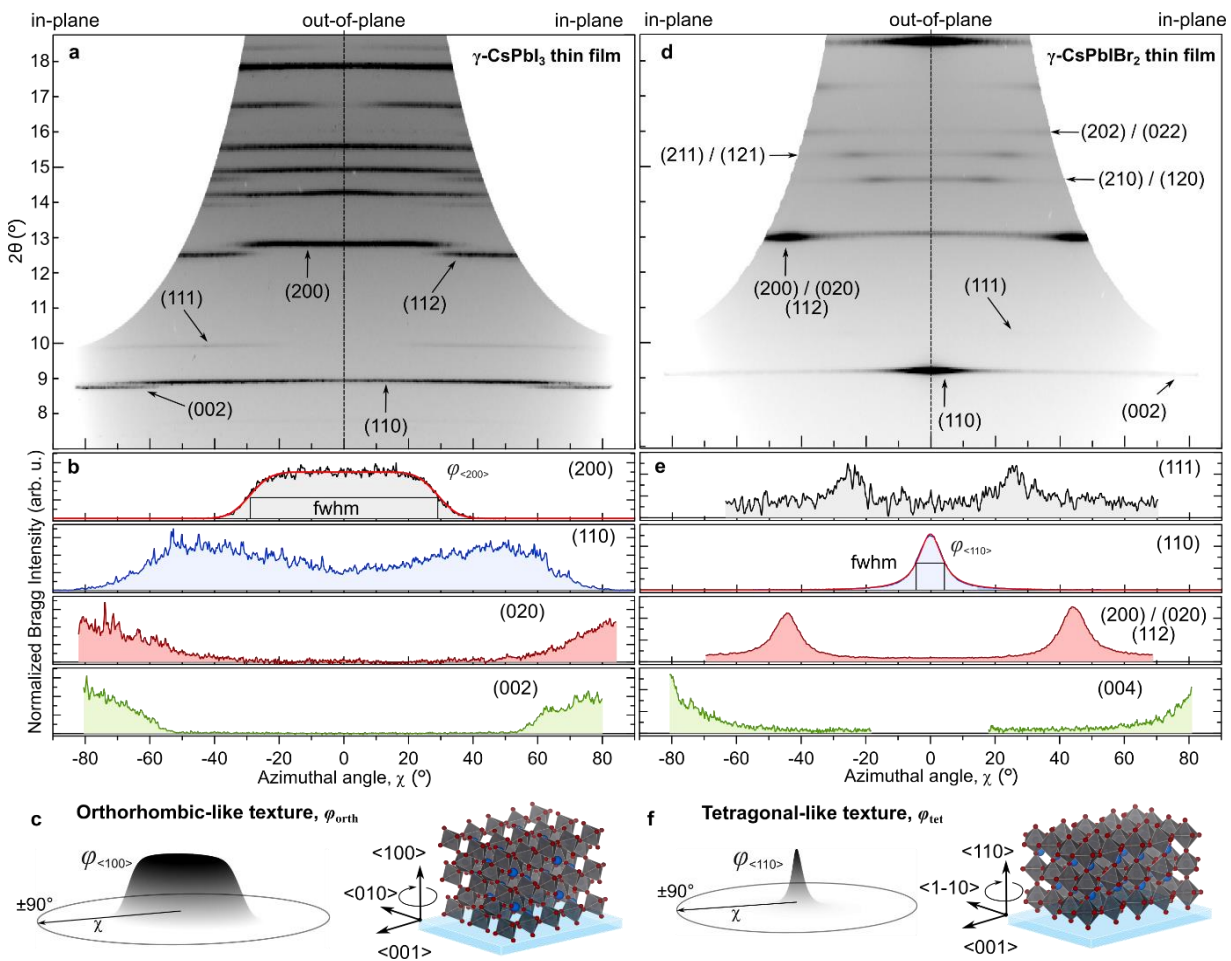

**Figure 5.** 2D GIWAXS intensity images (λ = 0.9577 Å) of (a) γ-CsPbI$_3$ and (d) γ-CsPbIBr$_2$ thin films (~300 nm thick) on glass substrates, as a linear function of the azimuthal angle. Both materials were annealed at 605 K for 1 minute before thermally quenching to an RT black phase. To extend the range of in-plane scattering signals which can be evaluated beyond what shown in (a) and (d) (i.e., with the incident beam centered on the detector), GIWAXS images were also recorded with the detector moved off-center (Figure S9 and S13). (b) and (e) display the corresponding normalized intensity profiles of different Bragg peaks as a function of the azimuthal angle. The red fit to the intensity distribution of (200) in (b), $\varphi_{<100>}$, is made using 2 CDF functions centered normal to the film plane (χ = 0°). The red fit made to the (110) distribution in (e) is a Lorentzian line shape. Geometric illustrations of (c) orthorhombic-like crystal texture, $\varphi_{orth.}$, and (f) tetragonal-like crystal texture, $\varphi_{tet.}$. Here their corresponding out-of-plane probability distributions for the thermally quenched γ-CsPbI$_3$ and γ-CsPbIBr$_2$ thin films are also depicted.

Br incorporation in γ-CsPbI$_{3-x}$Br$_x$ leads to a substantial reduction in the $e_{orth}$ strain component and a subsistence of $e_{tet}$ (Figure 4E and F). Thus, one can expect that Br-introduction can modify the polycrystalline texture formed in Br-rich CsPbI$_{3-x}$Br$_x$ thin films. Figure 5D presents a 2D GIWAXS intensity image of a thermally quenched γ-CsPbIBr$_2$ thin film with the Bragg intensities in the azimuthal domain. Off-center images, showing the (004) peak, are shown in Figure S13. The texture exhibited in thin film γ-CsPbIBr$_2$ is different to the pure triiodide system, having <110> normal to the planar surface in a relatively narrow distribution (*fwhm* of Lorentzian is ~10°). Consequently, the uniaxial texture limits the <1-10> and <001> directions to be in-plane. This results in the (200) and (020) lattice planes – which are approximately equal – to appear in a bi-modal fashion at an azimuthal angle of ± 45°. Likewise, the (112) peak in a tetragonal-like system arises close to the same 2θ and χ scattering angles, which is why it remains coupled in the corresponding intensity plot in Figure 5E. This texture formation is identified as tetragonal-like, $\varphi_{tet}$, and is displayed in Figure 5F. Tetragonal-like texture has been widely reported within hybrid organic-inorganic lead halide perovskite thin films[14,20,41] which tend to adopt tetragonal or tetragonal-like perovskite crystal structures. Moreover, a $\varphi_{tet}$ texture was recently reported in tetragonal β-CsPbI$_3$ thin films, indicating that the crystal symmetry, rather than the composition[13], is most influential.

In line with its crystal symmetry-driven origins, the direction and distribution of $\varphi_{orth}$ is found to remain steady across a broad range of annealing temperatures (Figure S14). Conversely, an analysis of γ-CsPbI$_2$Br thin films with an intermediate Br content reveal a hybrid texture formation (Figure S15:

simultaneous contributions of both $\varphi_{\text{orth}}$ and $\varphi_{\text{tet}}$). However, this dual texture is strongly dependent on the annealing temperature; higher annealing temperatures promote a relatively stronger $\varphi_{\text{tet}}$ expression and narrow its angular distribution (Figure S16). Similar dual textures have been reported previously in $CsPbI_{1.8}Br_{1.2}$ solar cells[42].

The bi-modal nature of the texture formation in quenched, black $CsPbI_{3-x}Br_x$ thin films, and its connection to crystal symmetry, suggest that these directions represent fundamental, energetically favored formations. For the case of γ-$CsPbI_3$ thin-films, anisotropic thermal expansion[31] arises because the tilting distortions are different. Here the normalized $b$- and $c$-axes of the γ-phase $CsPbI_3$ experience the smallest lattice reduction at RT, while the $a$-axis contracts much faster and ends up the smallest. We suggest that the introduction of tensile in-plane strain after quenching favors distortions that approximately align the $b$- and $c$-axes in-plane – directing the $a$-axis out-of-plane – to minimize the tensile strain introduced after cooling.

Next, we elucidate how strain-driven energy differences might arise in thermally treated $CsPbI_{3-x}Br_x$ thin films using periodic density functional theory (DFT; see Methods and Figure S17) to calculate the ground state energy difference of α-, β- and γ-phase $CsPbI_3$ perovskite. Applying ~1% strain along different crystal planes (Figure S18 and Table S1) shows that straining the (001) plane in both the β and γ phases is energetically unfavoured, and the c-axis should therefore preferentially align parallel to the interface, in agreement with experiment. Further, the length of the free lattice vector is always smaller than the strained lattice vectors (Table S1), with the shortest lattice vector oriented perpendicular to the strain. Additionally, biaxially straining (100) will be preferred over the (111) plane for the β phase and the (112) and (010) planes for the γ phase. This supports the notion that the longer $b$-and $c$-axes of the primitive β and γ unit cells will orient parallel to the interface. While this agrees directly with the experimental orthorhombic-like texture for γ-$CsPbI_3$ (Figure 5C), it is unexpected for the tetragonal-like texture of γ-$CsPbIBr_2$ (Figure 5F).

To understand this difference, it is important to investigate the ground state energy difference between the (100) and the (110) planes for both β- and the γ-$CsPbI_3$, taking into account that only the pure iodide material was theoretically investigated (Figure S18). Starting in the γ phase, Figure S18 demonstrates that biaxially straining the (110) plane in $CsPbI_3$ is about 19 meV per formula unit (p.f.u.) less favourable than biaxially straining the (100) plane. By decreasing the magnitude of the orthorhombic strain $e_{orth}$, this difference in ground state energy decreases to about 11 meV p.f.u. for the β phase of $CsPbI_3$. This observation can be explained given that the *a*-axis is significantly shorter than the *b*-axis for the γ phase while they are equal for the β-phase. The less pronounced preference for the (100) plane upon converting from the γ- to the β-phase in $CsPbI_3$ can therefore be correlated with the reduction in magnitude of $e_{orth}$. While biaxially straining (110) never becomes more favourable than straining the (100) plane for $CsPbI_3$ – even without orthorhombic strain an 11 meV p.f.u. difference remains – this clearly shows that reducing the orthorhombic strain is crucial to increase the likelihood of observing the tetragonal texture. Given that Figure 4E experimentally demonstrates that the magnitude of the orthorhombic strain decreases with increasing Br incorporation, this observation forms a possible explanation of why a tetragonal-like structure is obtained in tetragonal-like γ-$CsPbIBr_2$. Further, this hints that the total energy will depend on a thermodynamic surface term[34,43], as a high-symmetry plane like (110) is a preferred interface plane while in the perovskite phase. Within this scenario, it will be a trade-off between the preferential alignment due to biaxial strain and the interfacing surface that determines the observed texture.

In champion halide perovskites solar cells, the use of mesoporous titanium dioxide (m-$TiO_2$) substrates is commonplace,[44] which has, by design, a relatively rough surface morphology (Figure S11). We evaluated the texture formation in device-ready FTO/c-$TiO_2$/m-$TiO_2$/$CsPbI_{3-x}Br_x$ thin films over a solar-friendly compositional range; $0 \leq x \leq 1.2$. Figure S19 focuses on the (110) peak evolution with rising Br, showing that the rough surface of m-$TiO_2$ introduces a disorderly, hybrid texture distribution of both $\varphi_{orth}$ and $\varphi_{tet}$ in $CsPbI_{3-x}Br_x$ thin films, which is quickly shifted toward $\varphi_{tet}$ by increasing the Br content. This suggests that the added surface roughness does not allow the interfacial forces at the junction to be uniform enough to clearly discriminate between different texture formations.

In summary, we have systematically studied texture formation within polycrystalline thin films of all-inorganic metal halide perovskite $CsPbI_{3-x}Br_x$ using synchrotron X-ray diffraction, structure-sensitive electron microscopy and DFT. First, through the examination of free-standing polycrystalline ensembles, shared nanograin orientations are shown to be an inherent feature within soft, polymorphic perovskites, being triggered by thermal-induced distortions within the unit cell. When expressed at scale within typical solution-processed thin film device architectures (i.e. atop transparent glass substrates), we discovered two distinct texture formations manifesting within $CsPbI_{3-x}Br_x$ thin films. I-rich films are shown to lead to orthorhombic-like texture (<100> out-of-plane; <010> and <001> in-plane), while Br-rich films form tetragonal-like texture (<110> out-of-plane; <1-10> and <001> in-plane). Following an exhaustive exploration of material processing parameters like substrate chemistry and roughness, film thickness and annealing temperature, the amount of Br incorporation in $\gamma$-$CsPbI_{3-x}Br_x$ thin films is revealed to be the central governing factor toward producing different texture formations, suggesting they are energetically favored formations. This is because Br incorporation modifies the $\gamma$-$CsPbI_{3-x}Br_x$ crystal structure by reducing the orthorhombic lattice distortion (making it more tetragonal-like), which changes the domain selection criterion coordinating texture formation at the thin film-level. Our findings help to unlock the interpretation of several anomalous reports of anisotropic thin-film behavior in polycrystalline perovskite thin films, typically derived using linear or single pixel X-ray scattering detection. Comparing our findings to wider research on hybrid organic-inorganic halide perovskites, like the popular hybrid methylammonium and formamidinium lead halide perovskites, the insights revealed here unify texture formation within thermally processed perovskite thin films. Being able to fine tune specific, energetically favored polycrystalline texture formations will ultimately lead to engineering device-ready materials. Large-area SXRD appears best suited to decipher the statistical information encoded in the fine nanoscale grains and polycrystalline thin film texture. Only with precise statistical data, like that depicted in Figure 5, can the influence of texture on carrier diffusion and transport within thin films to be accurately quantified or understood, leading to improved device performance.

**Supporting Information**
Supporting Information is available from the Wiley Online Library or from the author.

**Acknowledgements**
J.A.S., T.B., and S.M.J.R. acknowledge financial support from the Research Foundation - Flanders (FWO: grant No.'s 12Y7218N, V439819N and 12Y7221N (J.A.S), 1SC1319 (T.B.), and 12T3519N (S.M.J.R.)). J.A.S. and M.B.J.R acknowledge financial support from the KU Leuven Industrial Research

Fund (C3/19/046). J.H and M.B.J.R acknowledge financial support from the Research Foundation - Flanders (FWO) through research projects (FWO Grant No's G098319N and ZW15_09-GOH6316), from the Flemish government through long term structural funding Methusalem (CASAS2, Meth/15/04) and the KU Leuven Research Fund (C14/15/053 and C19/19/079). M.B.J.R acknowledges financial support from the KU Leuven Research Fund (C14/19/079). J.H., V.V.S. and M.B.J.R. thank the Flemish government for iBOF funding (PERsist). The computational resources (Stevin Supercomputer Infrastructure) and services used in this work were provided by the VSC (Flemish Supercomputer Center), funded by Ghent University, FWO and the Flemish Government – department EWI. V.V.S acknowledges the Research Board of Ghent University (BOF). The authors thank S. Lammar for providing the $NiO_x$ substrates used in the study and J.A.S. thanks Prof. Peidong Yang for his thoughtful scientific discussions.

Received: ((will be filled in by the editorial staff))
Revised: ((will be filled in by the editorial staff))
Published online: ((will be filled in by the editorial staff))

References

[1] L. Chouhan, S. Ghimire, C. Subrahmanyam, T. Miyasaka, V. Biju, *Chem. Soc. Rev.* **2020**, 10.1039.C9CS00848A.
[2] H. J. Snaith, *Journal of Physical Chemistry Letters* **2013**, *4*, 3623.
[3] Q. Jiang, Y. Zhao, X. Zhang, X. Yang, Y. Chen, Z. Chu, Q. Ye, X. Li, Z. Yin, J. You, *Nat. Photonics* **2019**, *13*, 460.
[4] National Renewable Energy Laboratory, **2020**, DOI https://www.nrel.gov/pv/cell-efficiency.html.
[5] H. Jin, E. Debroye, M. Keshavarz, I. G. Scheblykin, M. B. J. Roeffaers, J. Hofkens, J. A. Steele, *Mater. Horiz.* **2020**, *7*, 397.
[6] W. Nie, H. Tsai, R. Asadpour, J.-C. Blancon, A. J. Neukirch, G. Gupta, J. J. Crochet, M. Chhowalla, S. Tretiak, M. A. Alam, H.-L. Wang, A. D. Mohite, *Science* **2015**, *347*, 522.
[7] X. Ren, Z. Yang, D. Yang, X. Zhang, D. Cui, Y. Liu, Q. Wei, H. Fan, S. (Frank) Liu, *Nanoscale* **2016**, *8*, 3816.
[8] H. Zhou, Q. Chen, G. Li, S. Luo, T. -b. Song, H.-S. Duan, Z. Hong, J. You, Y. Liu, Y. Yang, *Science* **2014**, *345*, 542.
[9] F. Sahli, J. Werner, B. A. Kamino, M. Bräuninger, R. Monnard, B. Paviet-Salomon, L. Barraud, L. Ding, J. J. Diaz Leon, D. Sacchetto, G. Cattaneo, M. Despeisse, M. Boccard, S. Nicolay, Q. Jeangros, B. Niesen, C. Ballif, *Nature Mater* **2018**, *17*, 820.
[10] M. Long, T. Zhang, H. Zhu, G. Li, F. Wang, W. Guo, Y. Chai, W. Chen, Q. Li, K. S. Wong, J. Xu, K. Yan, *Nano Energy* **2017**, *33*, 485.
[11] C. Jiang, Y. Xie, R. R. Lunt, T. W. Hamann, P. Zhang, *ACS Omega* **2018**, *3*, 3522.
[12] F. Ji, S. Pang, L. Zhang, Y. Zong, G. Cui, N. P. Padture, Y. Zhou, *ACS Energy Lett.* **2017**, *2*, 2727.
[13] Y. Wang, M. I. Dar, L. K. Ono, T. Zhang, M. Kan, Y. Li, L. Zhang, X. Wang, Y. Yang, X. Gao, Y. Qi, M. Grätzel, Y. Zhao, *Science* **2019**, *365*, 591.
[14] B. J. Foley, S. Cuthriell, S. Yazdi, A. Z. Chen, S. M. Guthrie, X. Deng, G. Giri, S.-H. Lee, K. Xiao, B. Doughty, Y.-Z. Ma, J. J. Choi, *Nano Lett.* **2018**, *18*, 6271.
[15] T. A. S. Doherty, A. J. Winchester, S. Macpherson, D. N. Johnstone, V. Pareek, E. M. Tennyson, S. Kosar, F. U. Kosasih, M. Anaya, M. Abdi-Jalebi, Z. Andaji-Garmaroudi, E. L. Wong, J. Madéo, Y.-H. Chiang, J.-S. Park, Y.-K. Jung, C. E. Petoukhoff, G. Divitini, M. K. L. Man, C. Ducati, A. Walsh, P. A. Midgley, K. M. Dani, S. D. Stranks, *Nature* **2020**, *580*, 360.
[16] I. M. Hermes, A. Best, L. Winkelmann, J. Mars, S. M. Vorpahl, M. Mezger, L. Collins, H.-J. Butt, D. S. Ginger, K. Koynov, S. Weber, *Energy Environ. Sci.* **2020**, 10.1039.D0EE01016B.
[17] W. L. Tan, Y. Y. Choo, W. Huang, X. Jiao, J. Lu, Y.-B. Cheng, C. R. McNeill, *ACS Appl. Mater. Interfaces* **2019**, *11*, 39930.

[18] T. D. Siegler, Y. Zhang, A. Dolocan, L. C. Reimnitz, A. Torabi, M. K. Abney, J. Choi, G. Cossio, D. W. Houck, E. T. Yu, X. Li, T. B. Harvey, D. J. Milliron, B. A. Korgel, *ACS Appl. Energy Mater.* **2019**, *2*, 6087.
[19] G. Zheng, C. Zhu, J. Ma, X. Zhang, G. Tang, R. Li, Y. Chen, L. Li, J. Hu, J. Hong, Q. Chen, X. Gao, H. Zhou, *Nat Commun* **2018**, *9*, 2793.
[20] A. Z. Chen, B. J. Foley, J. H. Ma, M. R. Alpert, J. S. Niezgoda, J. J. Choi, *J. Mater. Chem. A* **2017**, *5*, 7796.
[21] S. K. Yadavalli, Y. Zhou, N. P. Padture, *ACS Energy Lett.* **2018**, *3*, 63.
[22] Y. Yang, S. Feng, M. Li, W. Xu, G. Yin, Z. Wang, B. Sun, X. Gao, *Sci Rep* **2017**, *7*, 46724.
[23] G. Madras, B. J. McCoy, *Journal of Crystal Growth* **2005**, *279*, 466.
[24] S. R. J., E. G. E., M. Laura, P. E. S., K. B. A., P. J. B., H. M. T., J. M. B., H. A. Abbas, M. D. T., S. H. J., *Advanced Energy Materials* **2016**, *6*, 1502458.
[25] J. A. Steele, H. Jin, I. Dovgaliuk, R. F. Berger, T. Braeckevelt, H. Yuan, C. Martin, E. Solano, K. Lejaeghere, S. M. J. Rogge, C. Notebaert, W. Vandezande, K. P. F. Janssen, B. Goderis, E. Debroye, Y.-K. Wang, Y. Dong, D. Ma, M. Saidaminov, H. Tan, Z. Lu, V. Dyadkin, D. Chernyshov, V. Van Speybroeck, E. H. Sargent, J. Hofkens, M. B. J. Roeffaers, *Science* **2019**, *365*, 679.
[26] Y. Rong, Y. Hu, A. Mei, H. Tan, M. I. Saidaminov, S. I. Seok, M. D. McGehee, E. H. Sargent, H. Han, *Science* **2018**, *361*, eaat8235.
[27] C. J. Howard, H. T. Stokes, *Acta Cryst.* **2005**, *A61*, 93.
[28] J. A. Steele, M. Lai, Y. Zhang, Z. Lin, J. Hofkens, M. B. J. Roeffaers, P. Yang, *Acc. Mater. Res.* **2020**, accountsmr.0c00009.
[29] S. D. Stranks, G. E. Eperon, G. Grancini, C. Menelaou, M. J. P. Alcocer, T. Leijtens, L. M. Herz, A. Petrozza, H. J. Snaith, *Science* **2013**, *342*, 341.
[30] Q. Dong, Y. Fang, Y. Shao, P. Mulligan, J. Qiu, L. Cao, J. Huang, *Science* **2015**, *347*, 967.
[31] A. Marronnier, G. Roma, S. Boyer-Richard, L. Pedesseau, J. M. Jancu, Y. Bonnassieux, C. Katan, C. C. Stoumpos, M. G. Kanatzidis, J. Even, *ACS Nano* **2018**, *12*, 3477.
[32] J. Zhao, Y. Deng, H. Wei, X. Zheng, Z. Yu, Y. Shao, J. E. Shield, J. Huang, *Sci. Adv.* **2017**, *3*, eaao5616.
[33] R. E. Beal, D. J. Slotcavage, T. Leijtens, A. R. Bowring, R. A. Belisle, W. H. Nguyen, G. F. Burkhard, E. T. Hoke, M. D. McGehee, *The Journal of Physical Chemistry Letters* **2016**, *7*, 746.
[34] B. Zhao, S.-F. Jin, S. Huang, N. Liu, J.-Y. Ma, D.-J. Xue, Q. Han, J. Ding, Q.-Q. Ge, Y. Feng, J.-S. Hu, *Journal of the American Chemical Society* **2018**, *140*, 11716.
[35] C. G. Bischak, M. Lai, Z. Fan, D. Lu, P. David, D. Dong, H. Chen, A. S. Etman, T. Lei, J. Sun, M. Grünwald, D. T. Limmer, P. Yang, N. S. Ginsberg, *Matter* **2020**, *3*, 534.
[36] S. Sharma, N. Weiden, A. Weiss, *Zeitschrift für Physikalische Chemie* **1992**, *175*, 63.
[37] M. A. Carpenter, E. K. H. Salje, A. Graeme-Barber, *ejm* **1998**, *10*, 621.
[38] A. Putnis, *Introduction to Mineral Sciences*, Cambridge University Press, Cambridge [England] ; New York, **1992**.
[39] V. Craciun, D. Craciun, X. Wang, T. J. Anderson, R. K. Singh, *Journal of Optoelectronics and Advanced Materials* **2003**, *5*, 401.
[40] J. E. Keem, J. M. Honig, *Selected Electrical and Thermal Properties of Undoped Nickel Oxide*, West Lafayette, IN, USA, **1978**.
[41] V. Trifiletti, N. Manfredi, A. Listorti, D. Altamura, C. Giannini, S. Colella, G. Gigli, A. Rizzo, *Adv. Mater. Interfaces* **2016**, *3*, 1600493.
[42] Q. Wang, J. A. Smith, D. Skroblin, J. A. Steele, C. M. Wolff, P. Caprioglio, M. Stolterfoht, H. Köbler, M. Li, S.-H. Turren-Cruz, C. Gollwitzer, D. Neher, A. Abate, *Sol. RRL* **2020**, *4*, 2000213.
[43] F. Yang, C. Wang, Y. Pan, X. Zhou, X. Kong, W. Ji, *Chinese Phys. B* **2019**, *28*, 056402.
[44] H.-S. Kim, N.-G. Park, *J. Phys. Chem. Lett.* **2014**, *5*, 2927.

Polycrystalline all-inorganic $CsPbI_{3-x}Br_x$ perovskite exhibits pervasive texture expressions when solution processed into thin film optical devices. Synchrotron-based large area X-ray scattering techniques provide insights which connect the final texture formation to the crystal symmetry of the halide perovskite, which can be tuned via halide mixing. Both I-rich and Br-rich materials each exhibit two different, energetically favored texture directions

Julian A. Steele, Eduardo Solano, Handong Jin, Vittal Prakasam, Tom Braeckevelt, Haifeng Yuan, Zhenni Lin, René de Kloe, Qiong Wang, Sven M. J. Rogge, Veronique Van Speybroeck, Dmitry Chernyshov, Johan Hofkens, and Maarten B. J. Roeffaers

**Texture Formation in Polycrystalline Thin Films of All-Inorganic Lead Halide Perovskite**

ToC Figure:

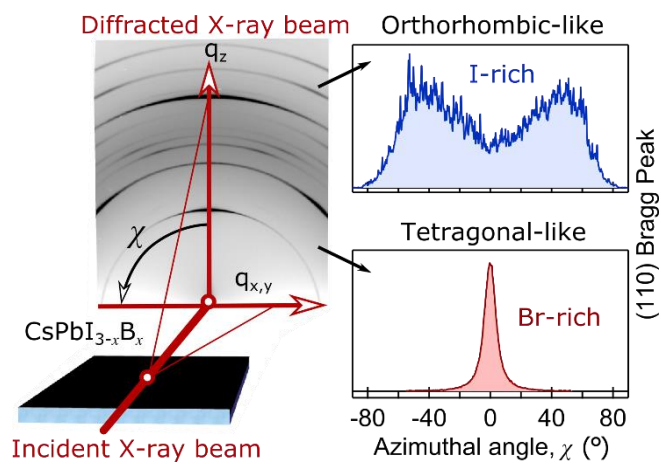

*Supporting Information for:*

# Texture Formation in Polycrystalline Thin Films of All-Inorganic Lead Halide Perovskite

Julian A. Steele,[1*] Eduardo Solano,[2] Handong Jin,[3] Vittal Prakasam,[1] Tom Braeckevelt,[3,4] Haifeng Yuan,[3] Zhenni Lin,[5,6] René de Kloe,[7] Qiong Wang,[8] Sven M. J. Rogge,[4] Veronique Van Speybroeck,[4] Dmitry Chernyshov,[9] Johan Hofkens,[3,10] and Maarten B. J. Roeffaers[1]

[1] cMACS, Department of Microbial and Molecular Systems, KU Leuven, 3001 Leuven, Belgium

[2] NCD-SWEET beamline, ALBA synchrotron light source, 08290, Cerdanyola del Vallès, Barcelona, Spain

[3] Department of Chemistry, KU Leuven, 3001 Heverlee, Belgium

[4] Center for Molecular Modeling (CMM), Ghent University, Technologiepark 46, 9052 Zwijnaarde, Belgium

[5] Department of Materials Science and Engineering, University of California, Berkeley, CA 94720, United states

[6] Materials Sciences Division, Lawrence Berkeley National Laboratory, Berkeley, CA 94720, United States

[7] EDAX, Ametek BV, Ringbaan Noord 103, 5046 AA Tilburg, The Netherlands

[8] Young Investigator Group Active Materials and Interfaces for Stable Perovskite Solar Cells, Helmholtz-Zentrum Berlin für Materialien und Energie, 12489 Berlin, Germany

[9] Swiss-Norwegian Beamlines at the European Synchrotron Radiation Facility, 71 Avenue des Martyrs, F-38043 Grenoble, France

[10] Max Plank Institute for Polymer Research, Mainz, D-55128, Germany

* julian.steele@kuleuven.be

# 1. Method

## A. Materials

**Perovskite Samples:** In order to totally remove water, CsI, PbI$_2$ and PbBr$_2$ powders were dried 12 h in vacuum oven at 60 ºC. A 0.2 mol/L perovskite precursor-solution is prepared in extra N,N-dimethylformamide (DMF) using CsI (Aldrich, 99.9%), PbI$_2$ (Aldrich, 99%) and PbBr$_2$ precursors mixed for the CsPbI$_{3-x}$Br$_x$ target composition in the appropriate molar ratio. The precursor-solution in DMF was filtered by a 0.45 µm PTFE filter and then either drop cast or spin coat onto a chosen substrate, in a nitrogen glove box. For either glass, glass/ITO or glass/ITO/NiO$_x$ substrates, the films were spin coat at 1500 rpm for 30 s and 3000 rpm for 60 s. In order to vary the thin film thickness, concentrations of 0.04, 0.1, 0.2 and 0.4 mol/L perovskite precursor-solutions were used. The higher concentration of 0.4 mol/L was used (at the solubility limit for the solution) with a spin-coating speed of 600 and 300 rpm for 30 s each, both followed with 1000 rpm for 60 s. All films were then transferred onto a hotplate for annealing at 160 ºC for 60 s. All films used in the main text were developed using a concentration of 0.2 mol/L, unless otherwise specified. The thin film thicknesses were determined using a 3D surface profilometry (S Lynx compact profilometer, Sensofar). CsPbBr$_3$ single crystals were prepared with a modified method from literature[1]. Sub-mm sized CsPbBr$_3$ crystals were prepared by heating an over-saturated precursor solution (with equal molar amounts of PbBr$_2$ and CsBr solids at the bottom of a 20 mL glass vial) on a hotplate at 75°C. Small crystals attaching to the sidewall of the glass vial can be collected after 48 h.

**Mesoporous TiO$_2$ substrates preparation:** Pre-patterned FTO substrates (TEC 15, Yingkou company, China) in the size of 2.5×2.5 cm$^2$ were cleaned via ultra-sonication, in 2% Mucasol solution (Schülke company), acetone and isopropanol for 15 minutes, respectively. After UV-ozone treatment, a TiO$_2$ compact layer was deposited on top of the substrates via spray pyrolysis, where the solution was composed of 0.48 ml acetyl acetone and 0.72 ml titanium diisopropoxide bis(acetylacetonate) in 10.8 ml of ethanol for 24 substrates. TiO$_2$ mesoporous layer was deposited from diluted TiO$_2$ paste (18NR, Great cell) in ethanol with the final solution density of ~ 0.8 mg/ml, and spin-coated at the speed of 4000 rpm for 10s, followed by annealing at 450°C for 30min. This results in a film thickness of roughly 150 nm, as measured by SEM.

**CsPbI$_{3-x}$Br$_x$ solar cell structures:** CsPbI$_{3-x}$Br$_x$ ($x$=1.2, 1.0, 0.8, 0.6, 0.3, and 0) inorganic perovskite were prepared from the mixed solution of PbI$_2$, PbBr$_2$ and CsI based on their desired stoichiometric ratios. More details can be found in our recent work.[2] The perovskite precursor was injected onto the substrate at 1000 rpm for 2 seconds and then 3000 rpm for 40 seconds, followed by annealing at 50 °C for $x$=1.2, 55 °C for $x$=1.0 and 0.8, and 60 °C for $x$=0.6, 0.3 and 0, then 100 °C for all samples, and then the final annealing temperature of 160 °C for $x$=1.2 and 1.0, 180 °C for $x$=0.8, and 310 °C for $x$=0.6 and 0.3, and 350 °C for $x$=0. Film thickness of around 500 nm to 550 nm was measured by SEM (Hitachi S-4100).

## B. Characterization

**Synchrotron X-ray Diffraction (SXRD):** SXRD data were collected at BM01 (SNBL/ESRF in Grenoble, France) using the PILATUS@SNBL diffractometer[3]. The monochromatic beam ($\lambda$ = 0.95774 Å) and the parameters of the detector were calibrated on LaB$_6$ powder using PyFAI[4]. The obtained calibrations were implemented to Bubble for further azimuthal integration of 2D images. The resulting unit cell models were refined using the Le Bail method in Fullprof[5].

CsPbI$_{2.85}$Br$_{0.15}$ powders were packed into a quartz capillary (200 µm diameter) under ambient conditions and placed in the synchrotron beam path (50 ×100 µm²; $\lambda$ = 0.957740 Å). The sample was positioned in parallel to the synchrotron beam and at an optimal point along the capillary to produce pronounced diffraction patterns. The position was fixed for further variable temperature measurements and the sample was not rotated. Temperature of the sample was regulated using the nitrogen hot blower, calibrated using the dependence of silver unit cell parameters, preliminarily measured at the sample position.

**Synchrotron-based Grazing Incidence Wide Angle X-ray Scattering (GIWAXS):** Data were recorded at NCD-SWEET beamline (ALBA synchrotron in Cerdanyola del Vallès, Spain) with a monochromatic ($\lambda = 0.95764$ Å) X-ray beam of $80 \times 30$ µm$^2$ [H × V], using a Si (111) channel cut monochromator. The scattered signal was recorded using a Rayonix LX255-HS area detector placed at 241.1 mm from the sample position. The reciprocal *q*-space and sample-to-detector distance were calculated using $Cr_2O_3$ as calibrant. An incident angle ($\alpha_i$) of 1° was chosen to ensure full penetration of the X-ray beam through the layer. Continuous $N_2$ flow over the sample was employed during the measurements. Collected 2D images were azimuthally integrated using PyFAI[4].

**Scanning Electron Microscopy (SEM):** SEM characterizations on the samples were carried out with a FEI Quanta 250 FEG SEM. An acceleration voltage of 2 or 5 kV was applied during the measurements to reduce charging on the sample surface. To avoid sample charging while imaging, a thin layer (<10 nm) of gold was deposited on the sample surface using vapour deposition.

**Electron back scatter diffraction (EBSD):** EBSD orientation measurements were obtained using a FEI Quanta 250 FEG SEM equipped with an EDAX Clarity direct electron EBSD detector. Patterns were collected using 15 kV accelerating voltage and 200 pA beam current. Detector exposure time was approximately 20 ms. The samples were uncoated to maximise the intensity of the diffraction signal. Due to the topology of the drop-cast crystals, full surface EBSD mapping was impossible because of shadowing artefacts. Instead, individual EBSD patterns were collected at the tips of the crystals in an aggregate. Where possible automatic band detection and indexing was applied. Otherwise the diffraction bands to use for indexing were marked manually.

**Atomic Force Microscopy (AFM):** The AFM images were obtained using Agilent Technologies 5100 AFM in ACAFM mode (tapping mode) using a Si tip (resonant frequency 100-300 kHz) at 512 point per line over a scan area of $8 \times 8$ µm$^2$.

**First principles study of texture formation:**

*General computational details.* Periodic first principles calculations were performed with the VASP software package[6,7] using density functional theory (DFT) with the SCAN (strongly constrained appropriately normed) exchange-correlation functional[8]. Convergence tests for the energy and forces were performed and afterwards the cut-off energy was set at 600 eV and a 3x3x3 k-points grid was chosen for the supercell. The considered valence electrons were $5s^2\ 5p^6\ 6s^1$ for Cs, $4d^{10}\ 5s^2\ 5p^2$ for Pb and $5s^2\ 5p^5$ for I. The ionic and electronic convergence threshold for the energy was set to $10^{-5}$ and $10^{-6}$ eV, respectively. The unstrained α, β and γ supercells were used as reference and were optimized with the conjugate gradient algorithm in VASP.

The biaxially strained α, β, and γ supercells (*vide infra*) were optimized using damped molecular dynamics in a user-modified VASP version. This modified version fixes the lengths of one or more lattice vectors along the principal axes of the cell during the optimization. To apply a biaxial strain in the (001) plane, for instance, the lengths of the ***a*** and ***b*** lattice vectors are set to the desired values and kept fixed during the optimization. For the (110) plane and similar planes not perpendicular to a principal axis, the unit cell first needs to be transformed to ensure the strained plane is one of the faces of the transformed cell. For instance, for the (110) plane, the new lattice vectors are ***a'*** = ***a*** + ***b***, ***b'*** = -***a***+***b***, and ***c'*** = ***c***. In this new unit cell, the same procedure as before can be followed by fixing the length of the ***b'*** and ***c'*** lattice vectors. To have a similar k-point density in these transformed cells, a 2x2x3 k-point grid is used, and the reference ground state energy for the unstrained cell has been recalculated with this new k-point density.

*Computationally applying a biaxial strain.* When decreasing the temperature after depositing an α-$CsPbI_3$ thin film on a glass substrate at 600 K, the much smaller thermal expansion coefficient of substrate compared to the perovskite and the strong attachment between the two materials at the interface hamper the shrinking of the interfacial plane of the α phase when decreasing the temperature. When reaching room temperature, this leads to biaxial strain at the interface of approximately 1%. To model this effect, 0 K geometric optimizations for the α, β, and γ phases of $CsPbI_3$ were performed in which the material was either unstrained or a strain was applied along a specific plane, representing the plane that forms the interface with the substrate. To this end, the following procedure was adopted.

First, the unstrained α phase structure was optimized. Then, the two most prominent low-index types of planes in the α phase were selected, *i.e.*, the (001) and the (101) planes. For each of these planes, the lattice dimensions were elongated with 1% and kept fixed (see Fig. S17) during a subsequent geometric optimization. The lattice dimensions of the biaxially strained (001) and (101) planes of the α supercell are respectively 12.74 Å x 12.74 Å and 12.74 Å x 18.02 Å. Afterwards, the new ground state energy for the strained cells was extracted.

For the β and γ phases, the same procedure was followed, although two aspects require specific attention in order to obtain consistent results. The first one is the lower symmetry of the β and γ phases. While all three principal directions of the cubic α phase are equivalent due to symmetry, this is not the case for the tetragonal β and orthorhombic γ phases. For instance, the (001), (010), and (100) planes are all equivalent for the α phase but no longer for the β and γ phases, requiring more planes to be investigated for these lower symmetry phases. For the β phase, the (100), (001), (110), and (111) planes are considered, while for the γ phase the (100), (001), (010), and (112) planes are considered. Note that, through the transformations outlined below and as visualised in Fig. S18, these planes all fall within the {001} and {101} families of planes of the α phase.

The second aspect requiring attention is the physically different composition of the (100) plane of the α phase compared to the (100) plane of the β and γ phases. For example, depending on the position of the (100) plane of the α phase, the plane will contain either Pb or Cs atoms but never both, while this can be the case for the (100) plane of the β phase. In fact, the composition of the (100) plane of the β phase resembles a lot better the (110) plane of the α phase. Therefore, it is physically not meaningful to compare the results obtained when straining the (100) plane of the α phase and the (100) plane of the β and γ phases. Ignoring the lower symmetry due to tilting of the octahedra, the unit cells of the β and γ phases can be obtained by transforming the primitive α phase following the relations $\boldsymbol{a}_\gamma = \boldsymbol{a}_\beta = \boldsymbol{a}_\alpha + \boldsymbol{b}_\alpha$, $\boldsymbol{b}_\gamma = \boldsymbol{b}_\beta = -\boldsymbol{a}_\alpha + \boldsymbol{b}_\alpha$, and $\boldsymbol{c}_\gamma = 2\boldsymbol{c}_\beta = 2\boldsymbol{c}_\alpha$. As the octahedra in the β and γ phases are tilted, making it impossible to map them back to the unit cell of the α phase, the three phases are mapped to a supercell, see Fig. S17. Using this transformation, planes defined for the similar supercells of three phases can be directly compared. Therefore, the biaxially strained lattice dimensions of the α supercell (12.74 Å x 12.74 Å and 12.74 Å x 18.02 Å for the (001) and (101) planes) can directly be used for the β and γ supercells.

## 2. Supporting Data

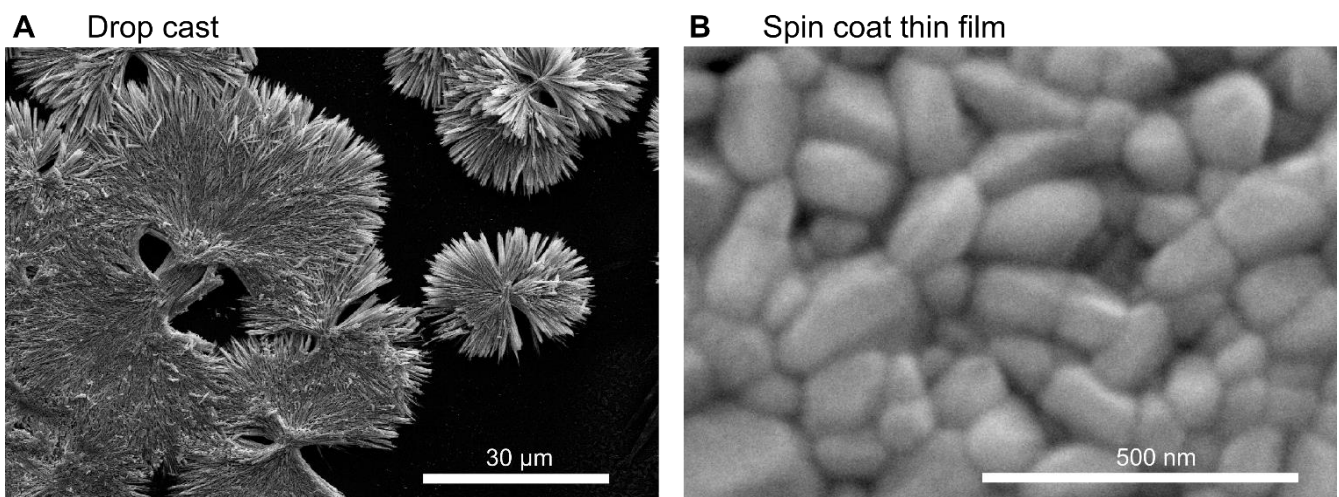

**Fig. S1:** Scanning electron microscope images of (A) polycrystalline $CsPbI_{2.85}Br_{0.15}$ samples made through drop cast deposition. These materials were then scraped from the substrate to form powders. (B) Spin coat polycrystalline $CsPbI_3$ thin film, showing representative grainy morphology of all-inorganic perovskite thin films.

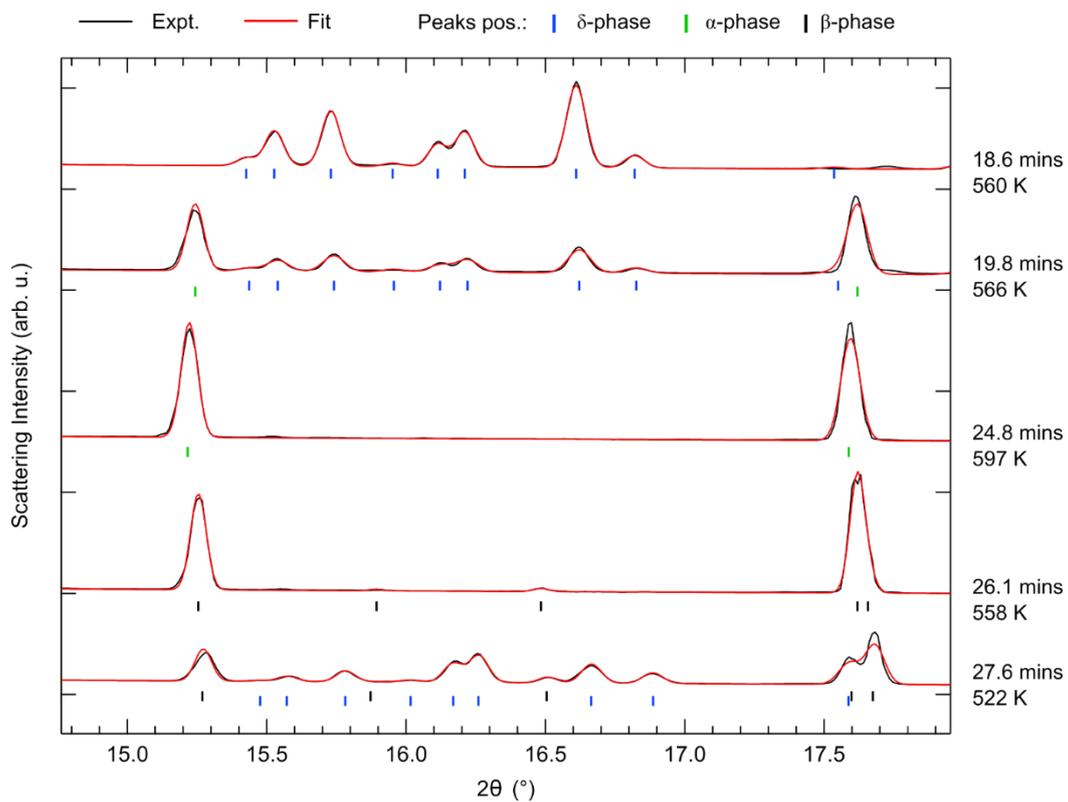

**Fig. S2:** SXRD patterns and their structural refinements (Le Bail method) of CsPbI$_{2.86}$Br$_{0.15}$ at different key points of the T-t profile shown in Figure 1A of the main text.

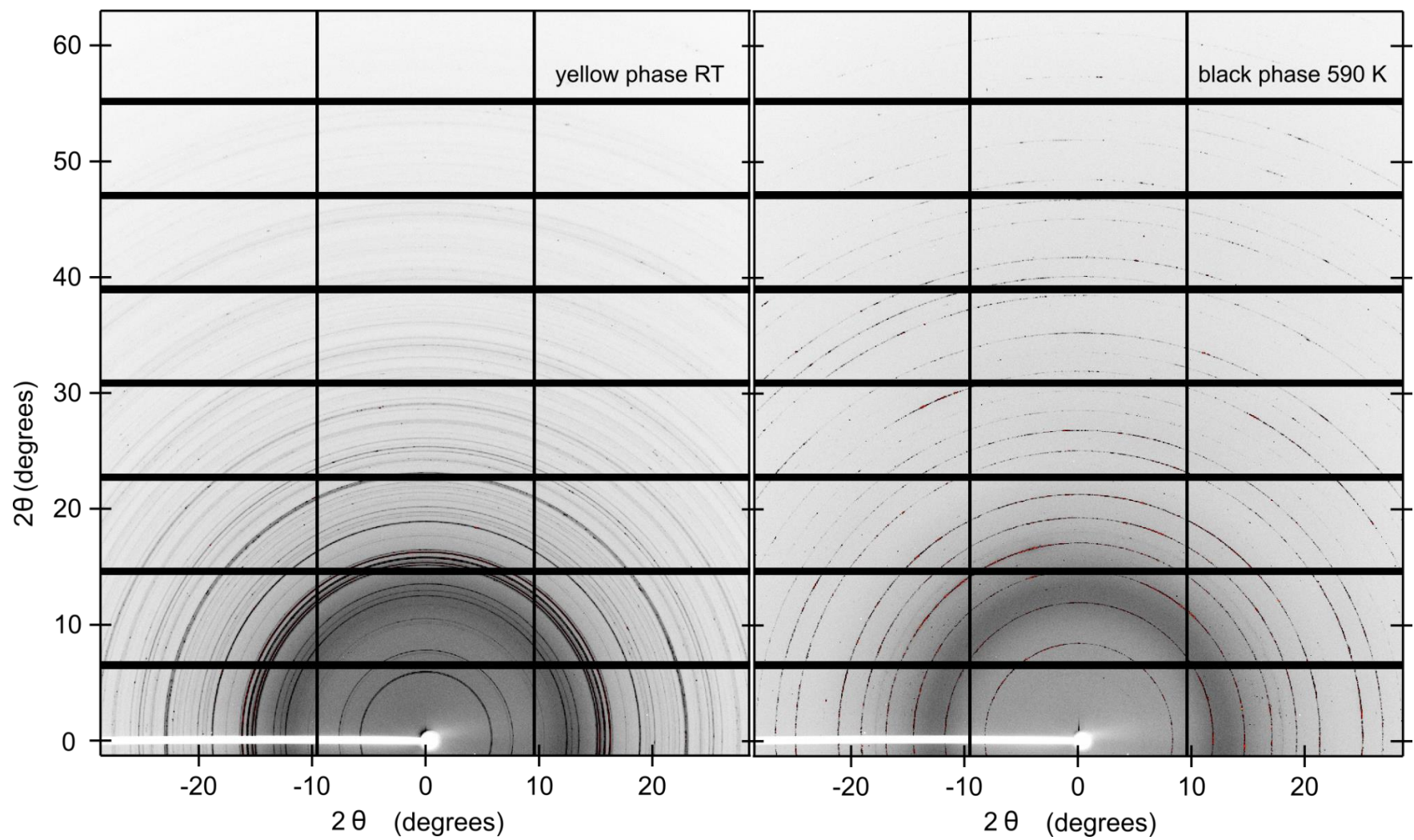

**Fig. S3:** Full 2D scattering patterns of data shown in Figure 1B of main text.

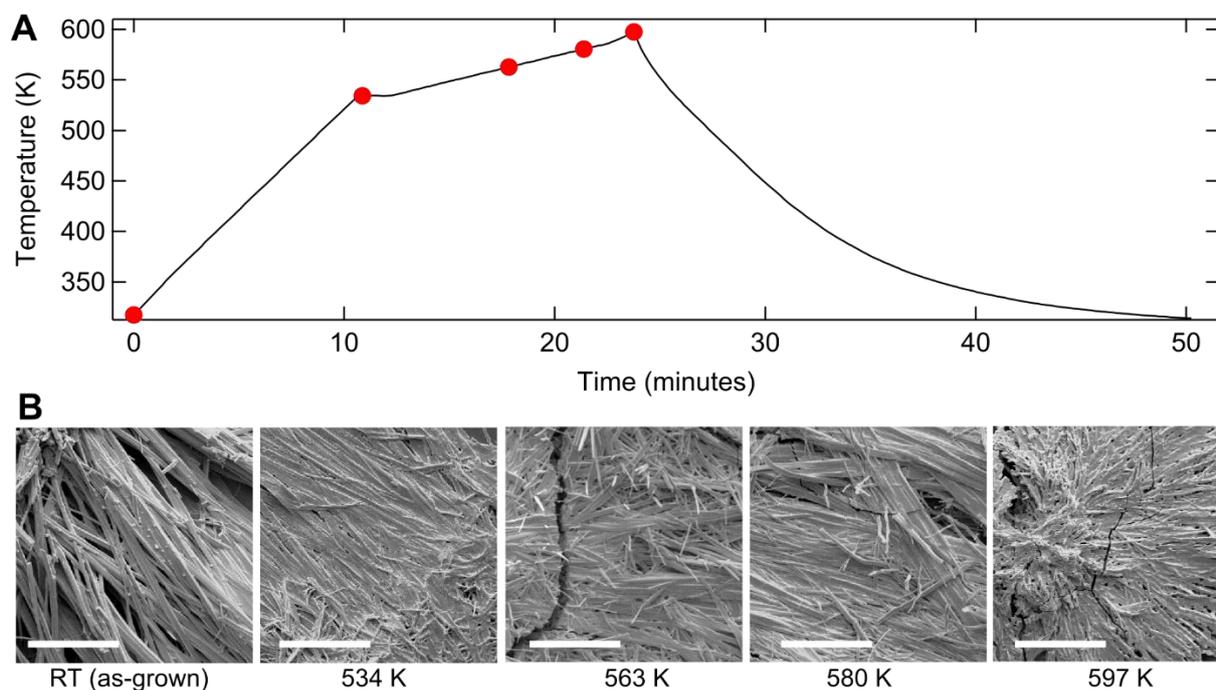

**Fig. S4:** (A) Full temperature-time profile of the heat treatment examined in Figure 1 of the main text. (B) Scanning electron microscope images of polycrystalline $CsPbI_{2.85}Br_{0.15}$ samples after reaching different high temperatures during the thermal ramp (red circles). Starting from stacks of needle-like crystals in the as-grown material, annealing causes widespread grains to coarsen and fuse. With large collection of grains joined, coherent stress fractures emerge in some after cooling. The scale bars represent a length of 30 µm.

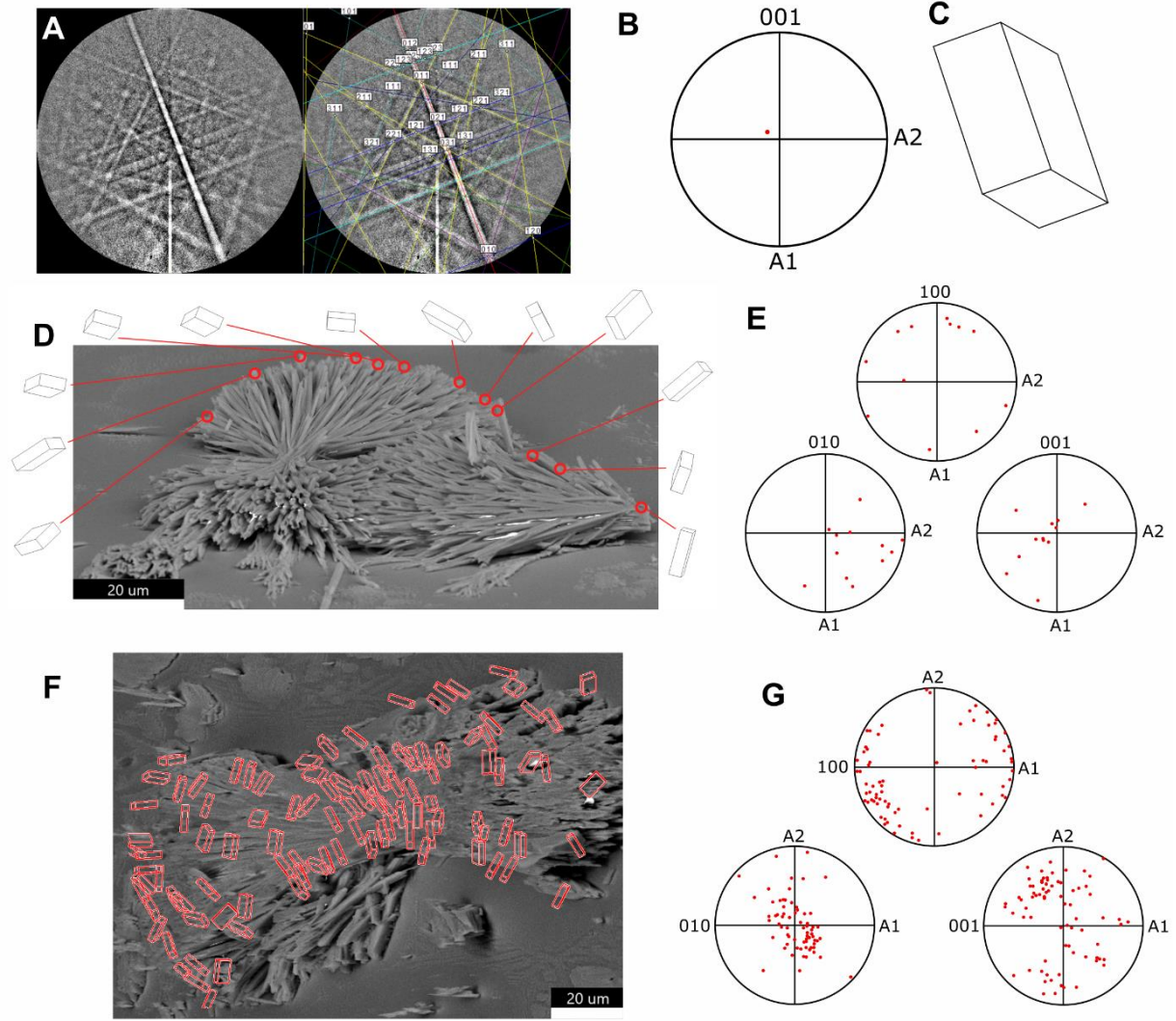

**Fig. S5:** (A) Left: EBSD pattern collected on a bulk crystal of $CsPbI_{2.85}Br_{0.15}$ following annealing. Right: Pattern with indexing solution overlay for the RT stable δ-phase. (B) (001) pole figure of the displayed orientation, and (C) wireframe illustration of the corresponding δ-phase unit cell orientation. (D) SEM image of as-grown δ-$CsPbI_{2.85}Br_{0.15}$ microcrystals with unit cells derived from the EBSD point measurements at the tips of 12 crystals. (E) Corresponding (100), (010), and (001) pole figures of the orientations of the crystals shown in (D). (F) SEM image of δ-$CsPbI_{2.85}Br_{0.15}$ microcrystals after annealing at 597 K, with wireframe orientations overlaid, as derived from EBSD point analyses. (G) Corresponding pole figures of the EBSD recorded from the crystals shown in (F). In both (D) and (F) the bundles of fibrous microcrystals can be recognised, but the individual fibres have been fused together in the case of the annealed sample. For the pole figures shown in (E) and (G), there is an intrinsic level of alignment between grains in the as-grown sample due to the restrictions imposed by the long, needle-like crystal morphology and the fact they are laying atop a planar substrate. This is especially apparent in the (100) pole figures where the points in the pole figure correspond with the direction of the needles. From these polar plots high temperature sintering is not seen to align crystals or form widespread grain growth. For the latter case, this suggests no coherence between neighbouring grains, which would result in much tighter concentration of orientations in the EBSD pole figures.

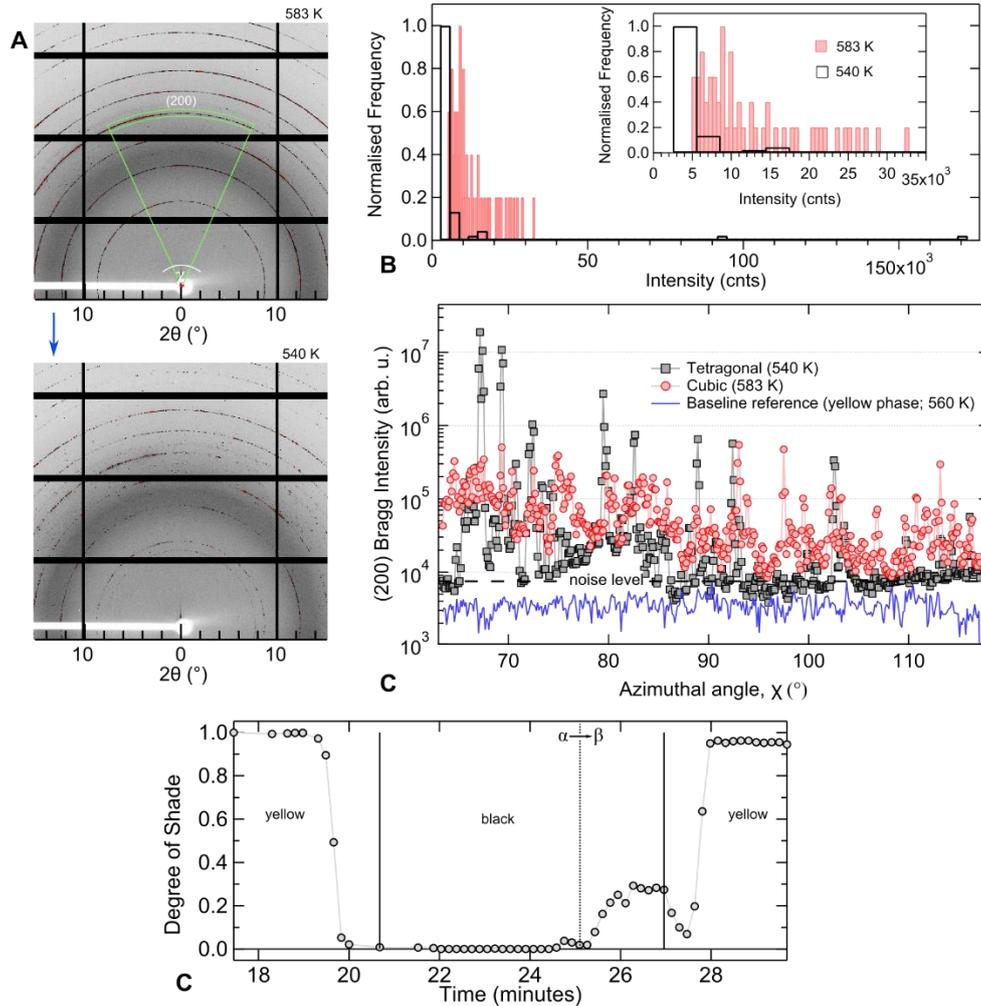

**Fig. S6:** (A) Examples of SXRD scattering image frames recorded in situ from the $CsPbI_{2.85}Br_{0.15}$ powdered sample in its high-temperature cubic phase (top) and its textured tetragonal phase during cooling (bottom). Similar frames were recorded throughout the thermal treatment presented in Figure 2(a) of the main text. Here in the top image, the green polar arc outlines the azimuthal region of the (200) Bragg peak used to analyse pixel shading and texture development within all recorded frames. We consider this arc as a set of pixels or azimuthal intervals and evaluate the intensity distribution across the arc. (B) compares the histogram distribution of the recorded intensities of the two frames in (A), highlighting the large shift is in the low-intensity signals in the inset. The skewed distribution of the tetragonal phase (540 K) toward low intensities indicates large shadowing across the evaluated arc. (C) Bragg intensity across the (200) diffraction arc in (A) as a function of the azimuthal angle, along with a comparison to a baseline signal recorded from the pure yellow phase. The dashed line shows the threshold level used to monitor the number of pixels that fall within the baseline noise range. The proportion of signals falling within the baseline noise represent the degree of texture-induced shading. (D) Degree of shading as a function of time across the production of a high-temperature black phase, separating the yellow phase at lower temperatures. The yellow phase will skew the shading analysis by reducing black phase scattering volume, which introduces a shading-like effect. Thus, frames containing detectable yellow phase signals are excluded from the t-T shading evolution analysis in Figure 3(a) of the main text.

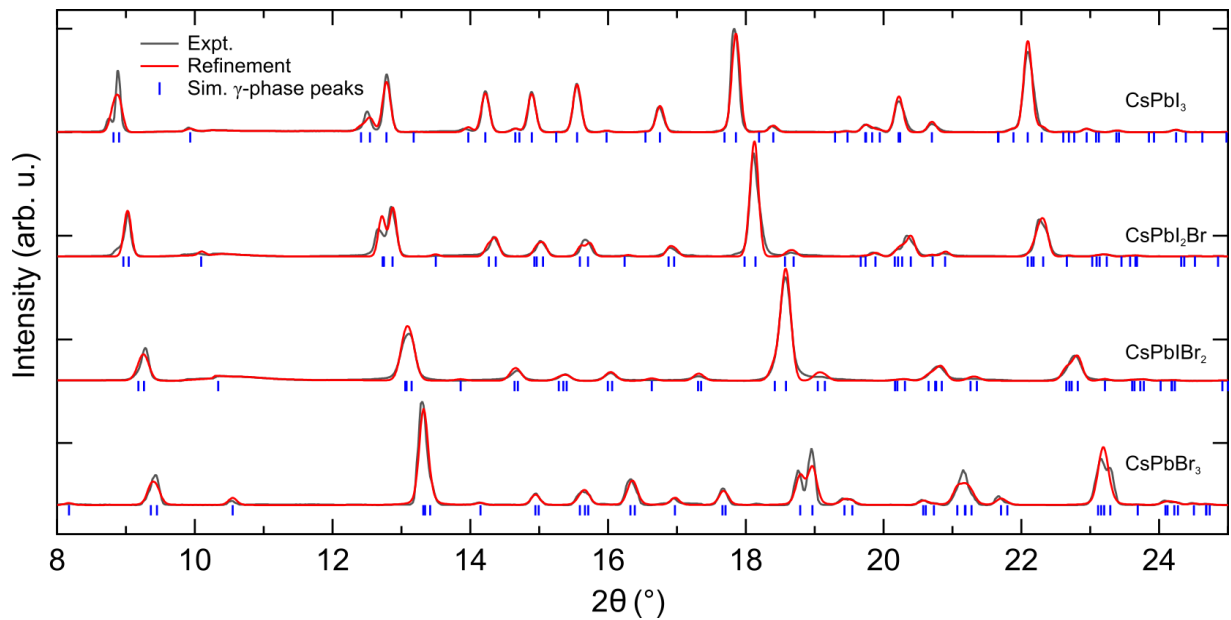

**Fig. S7:** Normalized SXRD patterns and their structural refinements (Le Bail method) of thermally quenched, black-phase $CsPbI_3$, $CsPbI_2Br$, and $CsPbIBr_2$ thin films, and thermodynamically stable γ-$CsPbBr_3$ single crystal. For clarity, these data have been background corrected and offset.

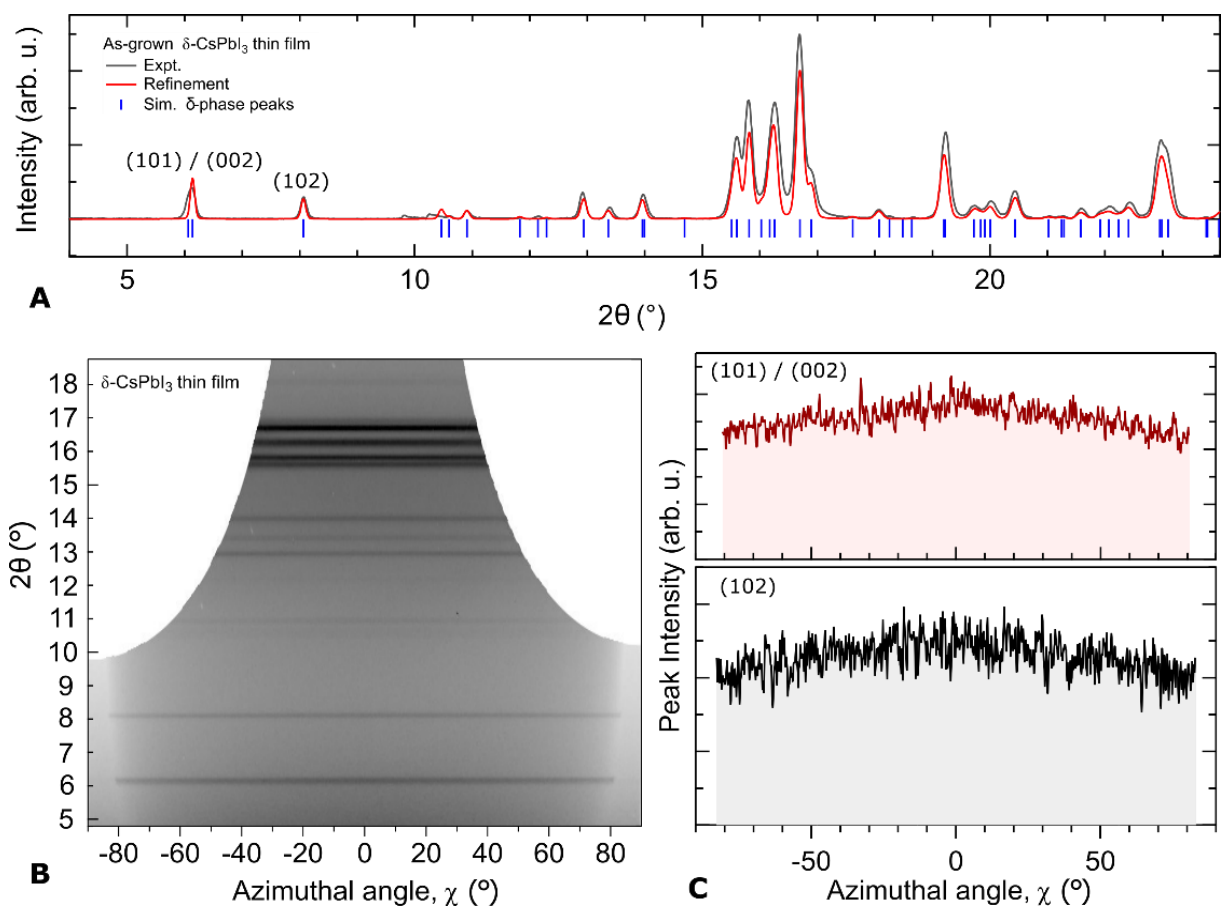

**Fig. S8:** (A) SXRD scattering pattern and structural refinement of as-grown CsPbI$_3$ thin film. (B) GIWAXS scan of as-grown thin film adapted to show scattering 2D intensity as a function of the azimuthal angle, χ. (C) Intensity of the low angle peaks as a function of the azimuthal angle, revealing no preferential orientation for these scattering planes, i.e. no texture.

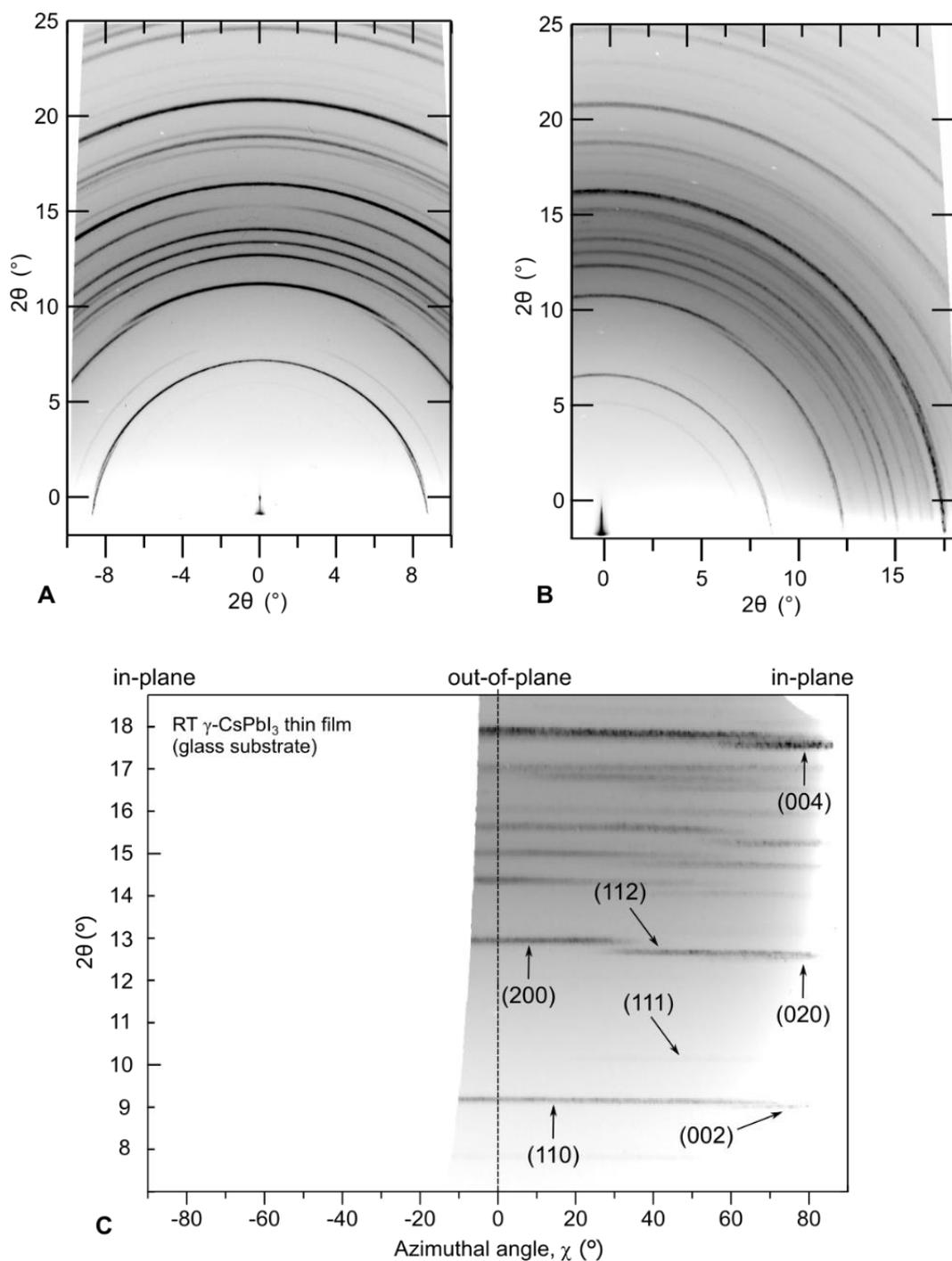

**Fig. S9:** 2D SXRD intensity images (λ = 0.9577 Å) of γ-CsPbI$_3$ thin film (~300 nm thick) on a glass substrate. (A) and (B) show the raw images recorded with the beam centre and off-centre, respectively. (C) Presents the off-centre image in (B) as a linear function of the azimuthal angle, with various peaks of interest identified.

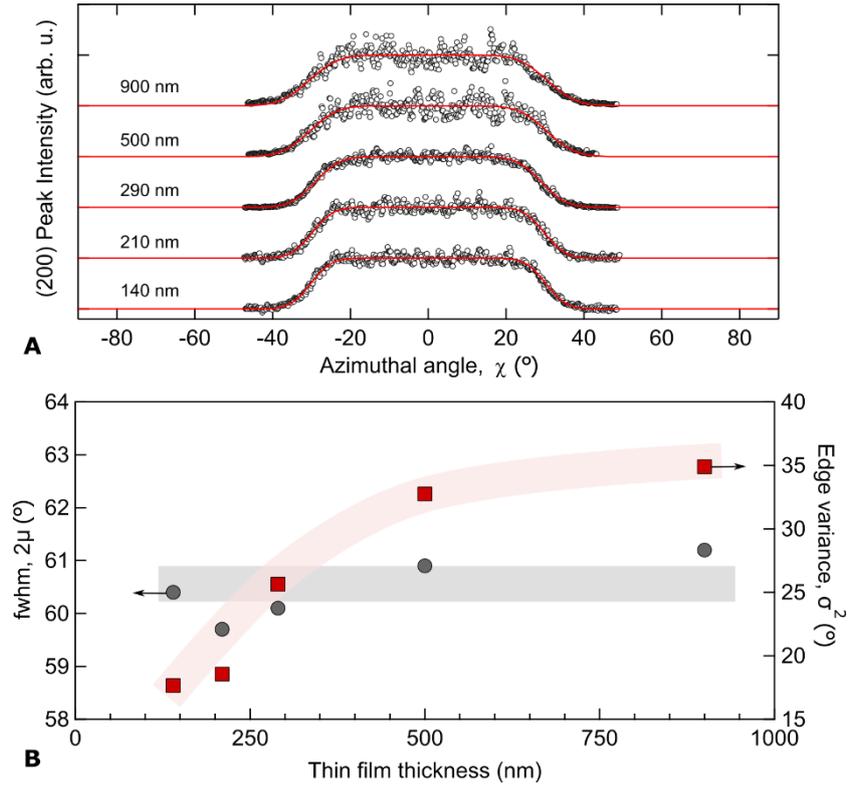

**Fig. S10:** Comparison of the (200) Bragg peak intensity distribution in the azimuthal domain for RT black CsPbI$_3$ thin films of different thicknesses. The fits (red) are made using a dual step function centred at χ = 0°: $\varphi_{<100>}(\chi) = A \times [1 - (\text{CDF}_+ + \text{CDF}_-)]$, where $\text{CDF}_\pm = \frac{1}{2}\left[1 + \text{erf}\left(\frac{\pm\chi-\mu}{\sigma\sqrt{2}}\right)\right]$. Here *A* is a constant, *erf* is the error function, $\mu$ describes the mean deviation from the substrate normal (2$\mu$ = fwhm), and $\sigma^2$ is the skew rate. (B) Fit parameters used to describe the (200) texture as a function of the thin film thickness. The thick lines are guides for the eye.

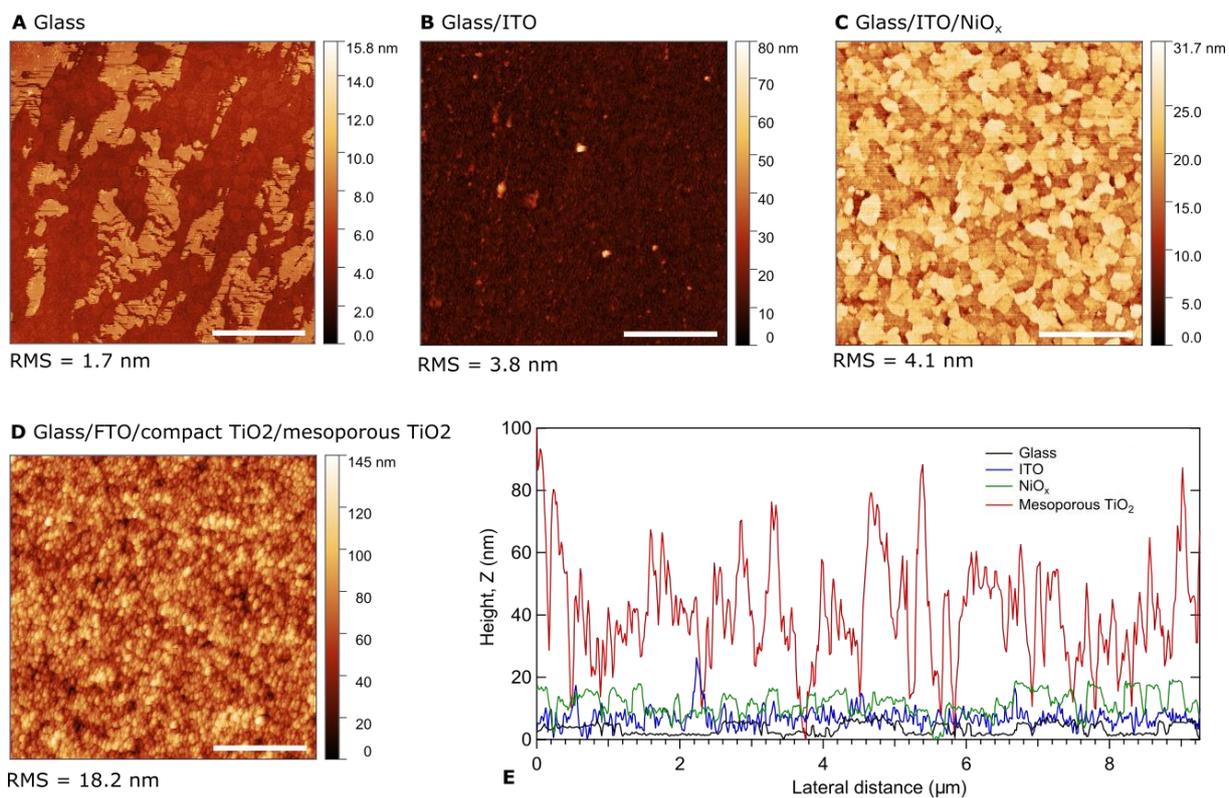

**Fig. S11:** Atomic force microscopy (AFM) maps of the different substrates and layers used in this study; (A) Glass, (B) Glass/ITO, (C) Glass/ITO/NiO$_x$ and (D) a mesoporous TiO$_2$ substrates, suitable for photovoltaics. The room mean squared (RMS) roughness values are provided at the bottom of each image. (E) Comparison of height line scans recorded from top left to the bottom right corners of each AFM image.

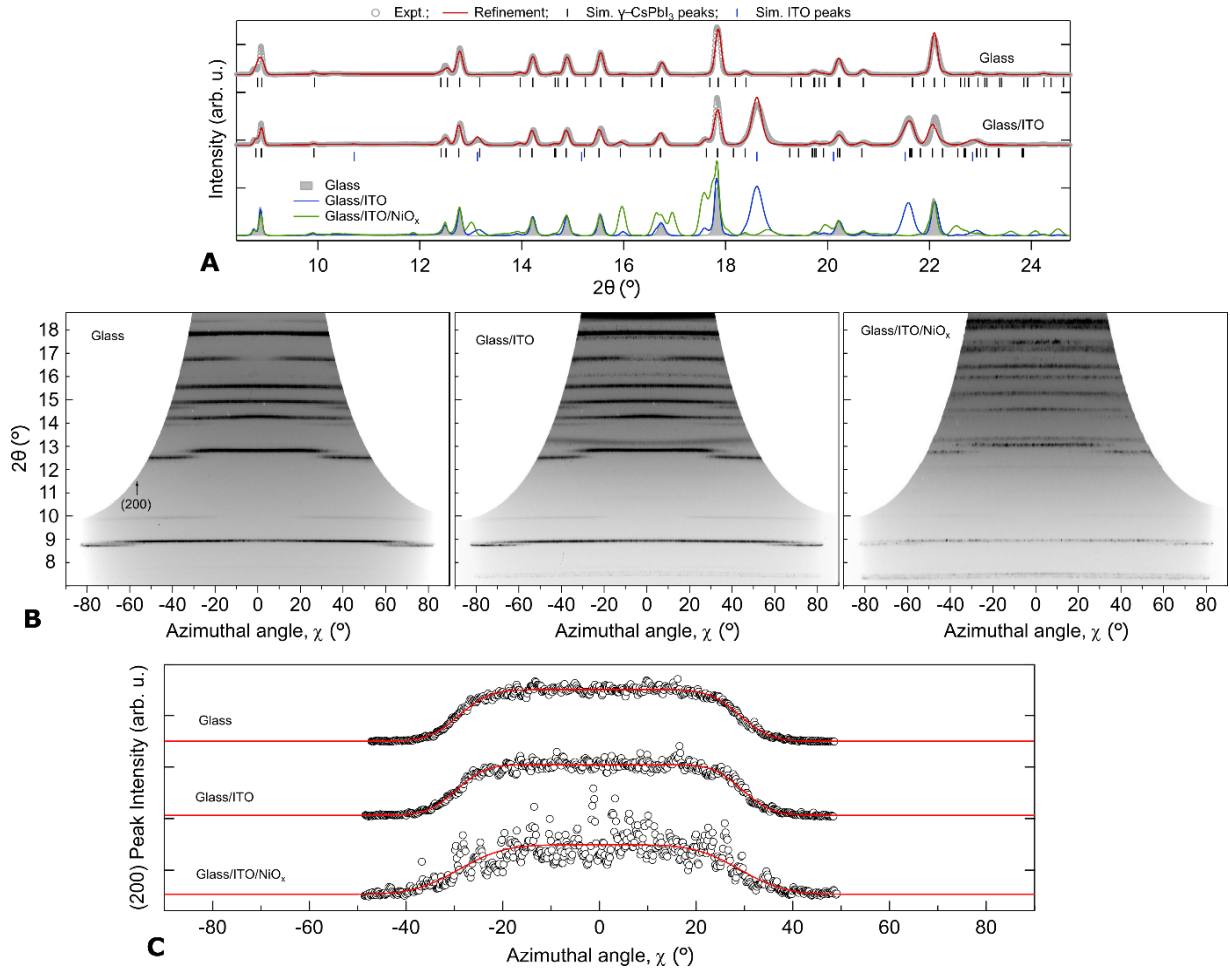

**Fig. S12:** (A) SXRD patterns of thermally quenched, black-phase CsPbI$_3$ thin films deposited onto different substrates. (B) GIWAXS images recorded from the thin films deposited into different substrates as a linear function of the azimuthal angle, with the (200) identified. (C) Comparison of the (200) Bragg peak intensity distribution in the azimuthal domain for RT black CsPbI$_3$ thin films deposited onto different substrates. The fits (red) are made using the dual step function outlined in the main text. For reference, the thermal expansion coefficient of the perovskite layer (~50×10$^{-6}$ K$^{-1}$) is relatively higher than glass (~4×10$^{-6}$ K$^{-1}$) or ITO[9] (~8×10$^{-6}$ K$^{-1}$) and NiO$_x$[10] (~10×10$^{-6}$ K$^{-1}$) layers, resulting in similar tensile biaxial strain in each case.

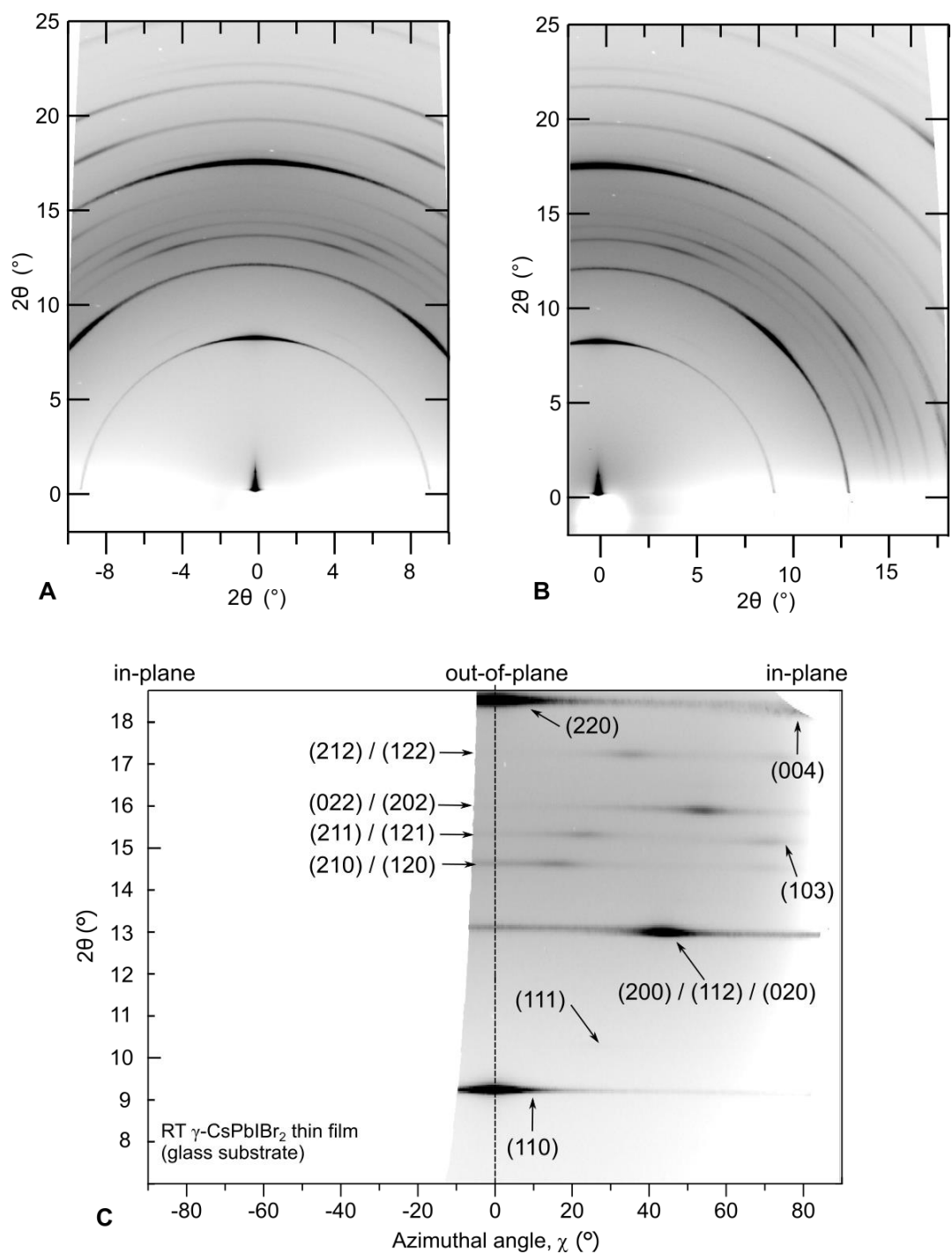

**Fig. S13:** 2D SXRD intensity images ($\lambda = 0.9577$ Å) of γ-CsPbIBr$_2$ thin film on a glass substrate. (A) and (B) show the raw images recorded with the beam centre and off-centre, respectively. (C) Presents the off-centre image in (B) as a linear function of the azimuthal angle, with various peaks of interest identified.

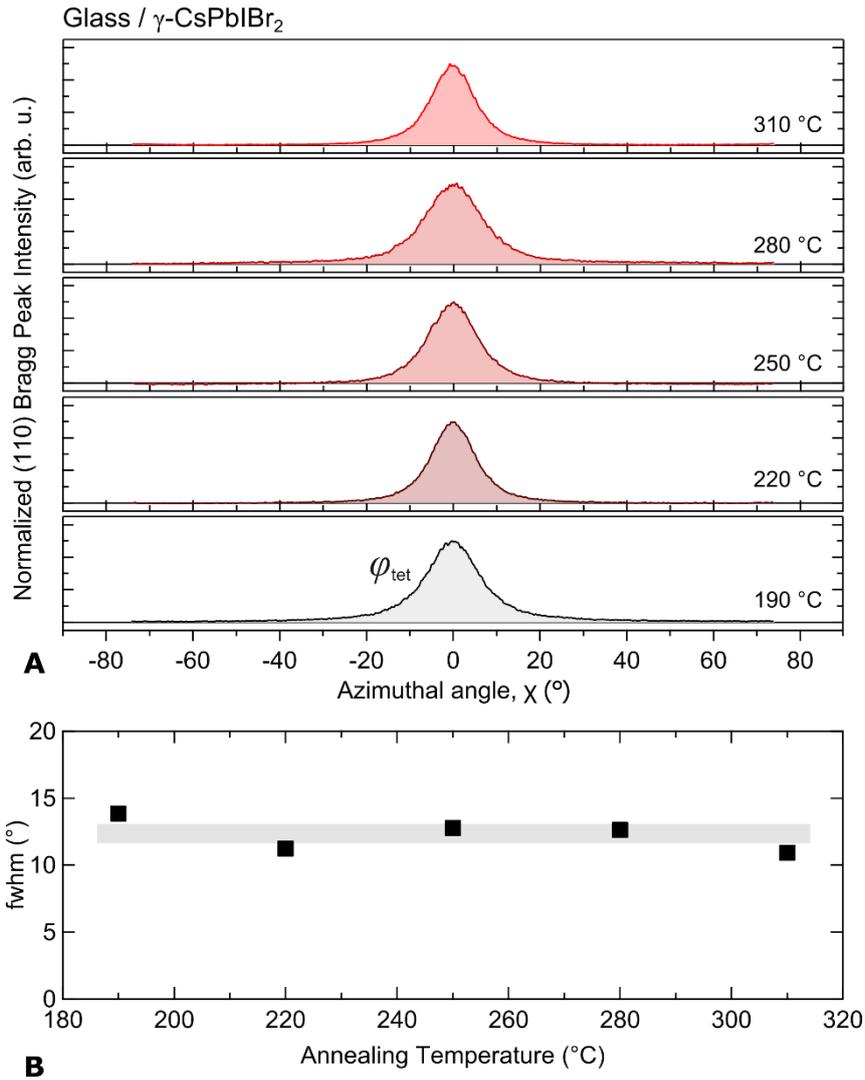

**Fig. S14: (A)** Comparison of the (110) Bragg intensity distribution recorded from CsPbIBr$_2$ thin film annealed at different temperatures for 1 minute. (B) The temperature-dependent full width at half maximum (fwhm) of the orthorhombic-like ($\varphi_{orth}$) texture which is described using a Lorentzian line shape. The solid grey line is a guide for the eye.

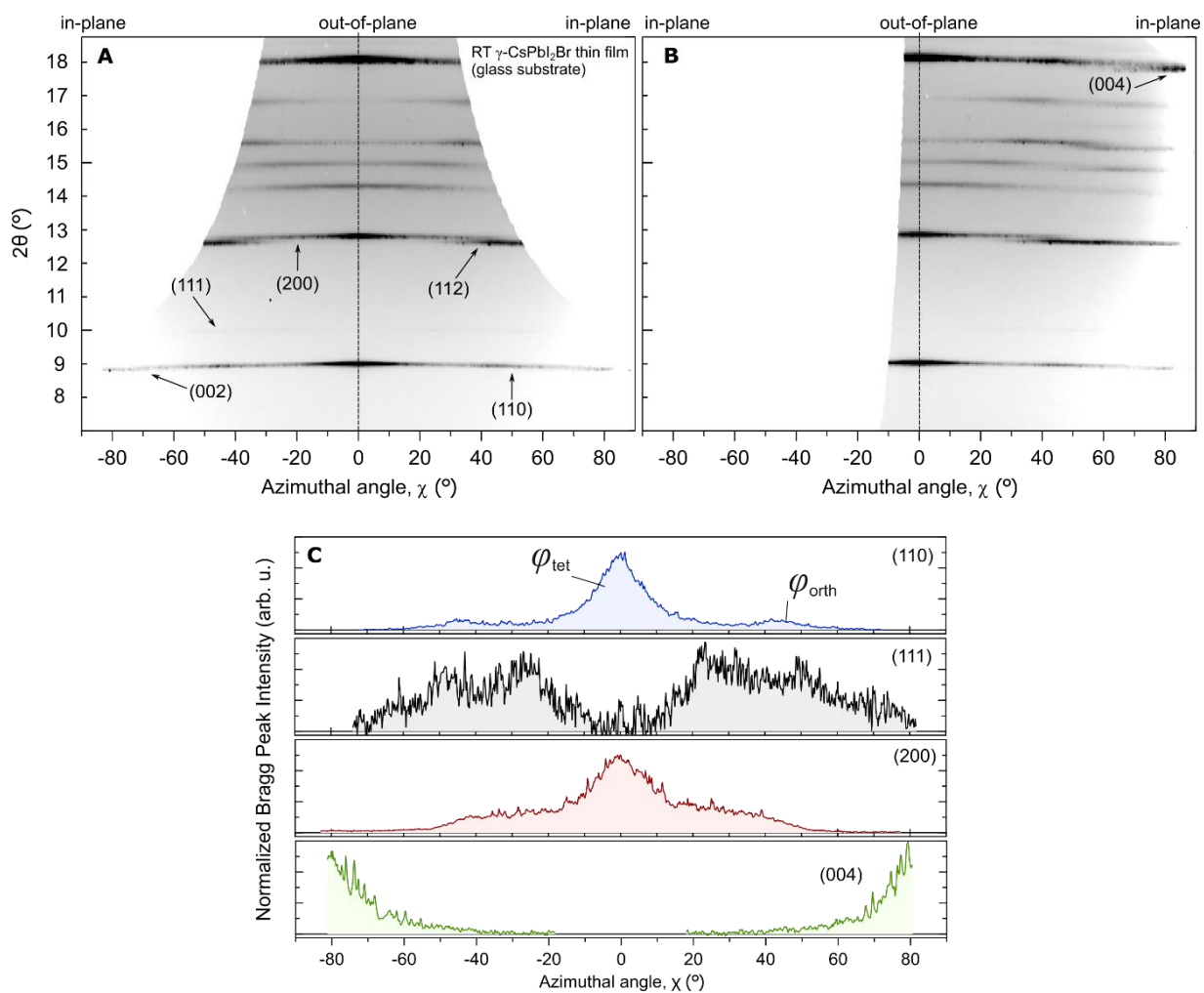

**Fig. S15:** CsPbI$_2$Br. 2D SXRD intensity images ($\lambda$ = 0.9577 Å) of γ-CsPbI$_2$Br thin film atop glass, recorded at the (A) centre and (B) edge of the large-area detector. The intensity images are presented as a linear function of the azimuthal angle. The thin film was thermally quenched into a black phase by annealing at 320°C for 1 minute before cooling quickly to RT black phase. (C) Corresponding normalized intensity profiles of various peaks found in the 2D SXRD images, as a function of the azimuthal angle. Co-existing orthorhombic-like ($\varphi_{orth.}$) and tetragonal-like crystal texture ($\varphi_{tet.}$) features are provided for the (110) distribution.

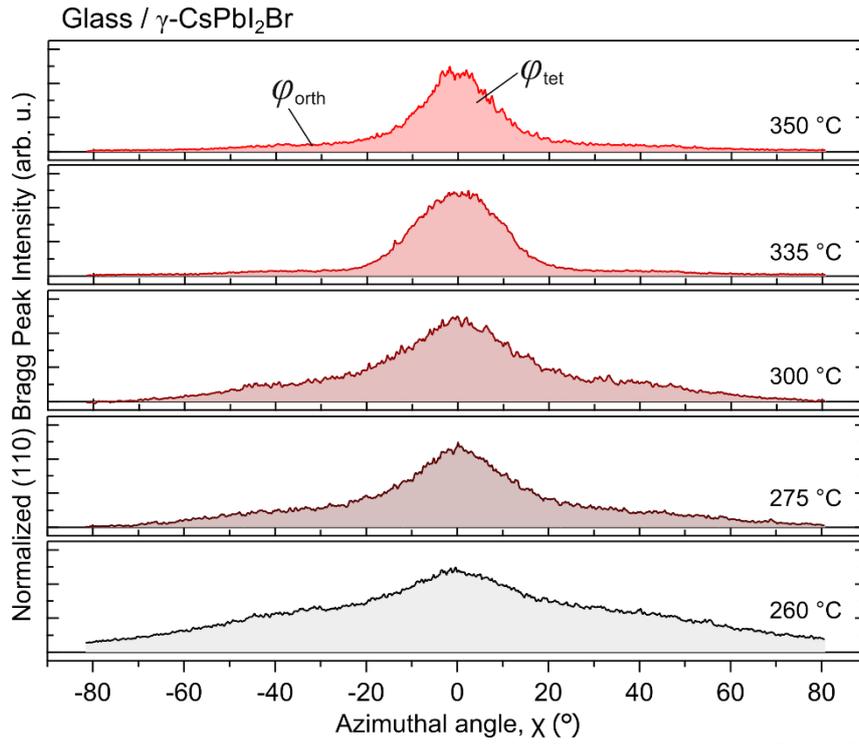

**Fig. S16:** Comparison of the (110) Bragg intensity distribution recorded from $CsPbI_2Br$ thin film annealed at different temperatures for 1 minute. Co-existing orthorhombic-like ($\varphi_{orth}$) and tetragonal-like crystal texture ($\varphi_{tet}$) features are identified and detailed in Figure 5 of the main text.

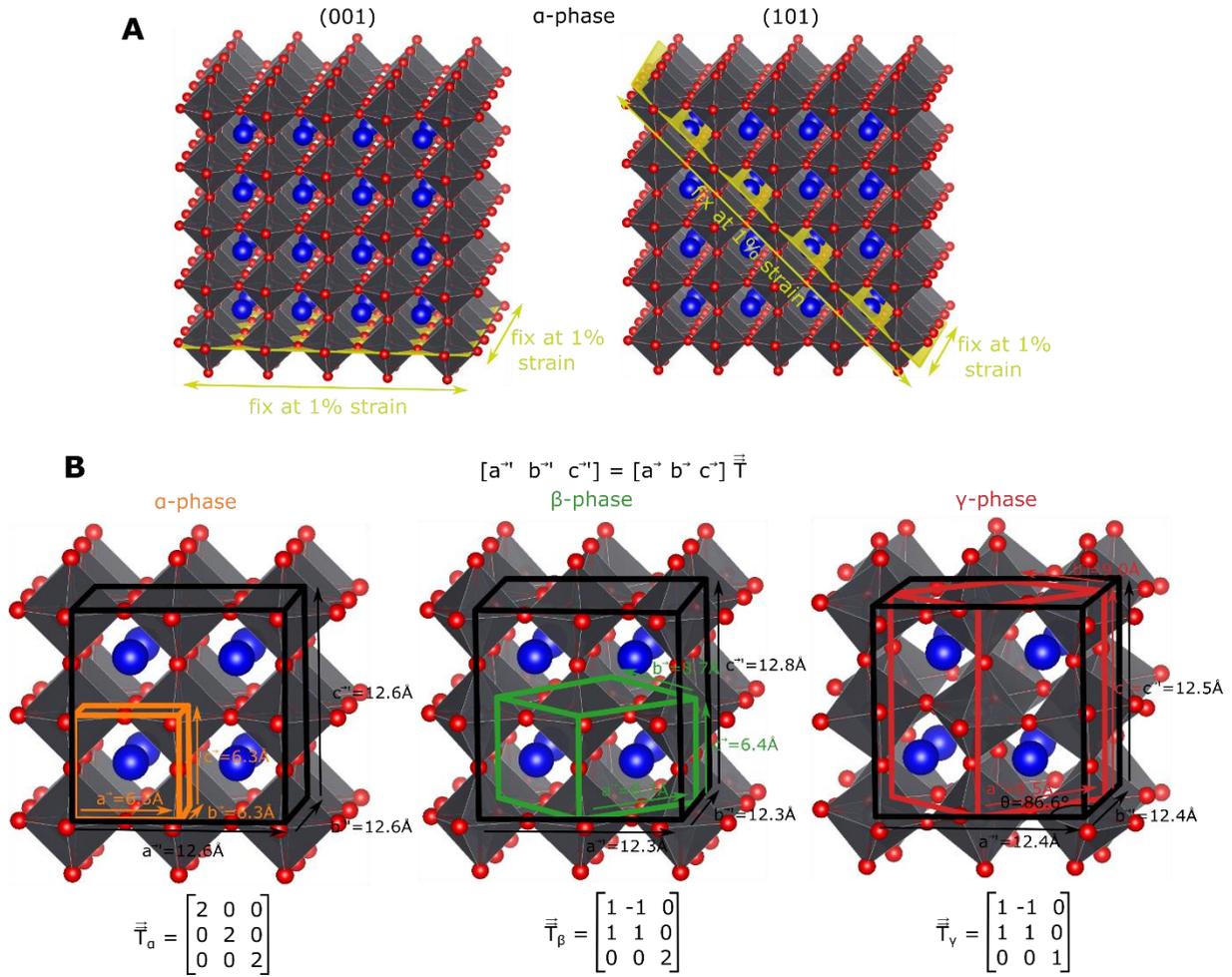

**Fig. S17.** (A) Indication of the (001) and (101) planes for the α phase of CsPbI$_3$. (B) Illustration of the α (orange), β (green) and γ (red) phases. On top of those figures their corresponding primitive unit cells and supercells are shown with the lengths of the lattice vectors and also a non-perpendicular unit cell angle of the γ supercell. The transformation matrix to transform the primitive unit cell to the supercell are shown below for all three phases.

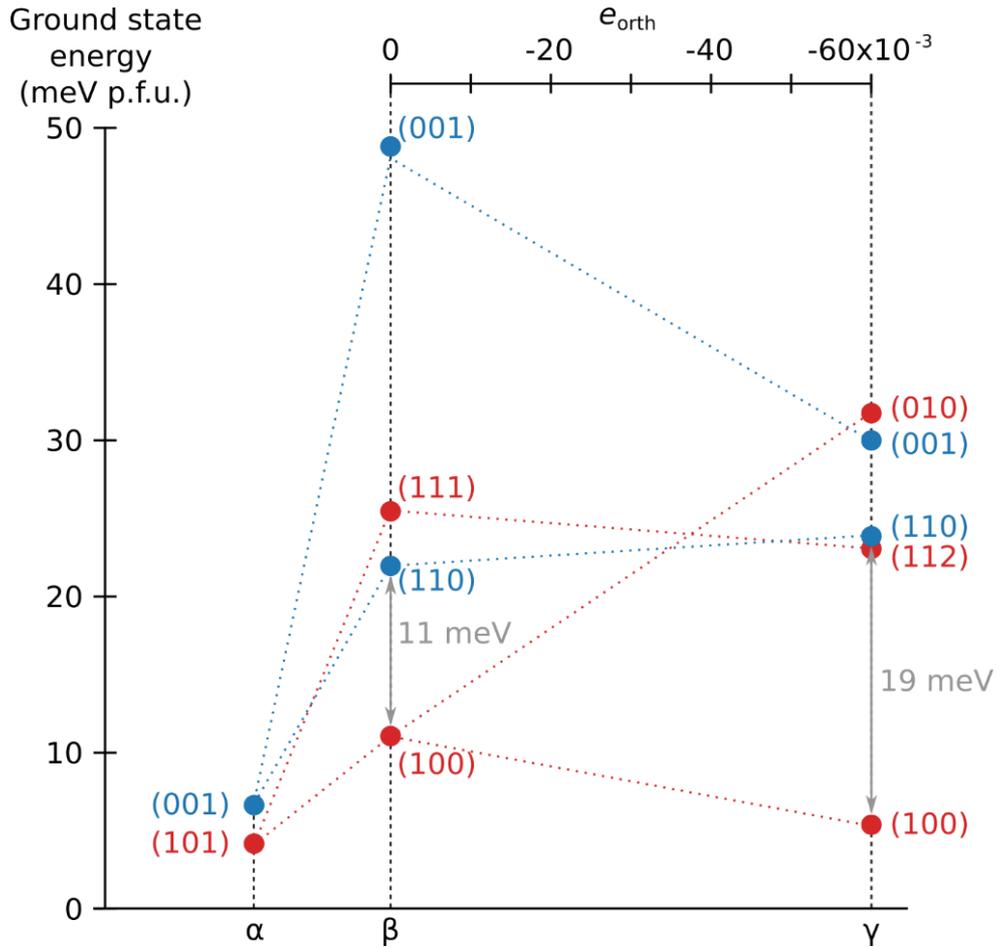

**Fig S18.** The ground state energy differences between the biaxially strained CsPbI$_3$ supercell along various planes and the optimized supercell for the α, β and γ phases. Blue and red planes correspond to the {001} and {101} families of planes in the cubic α phase. The orthorhombic ($e_{orth}$) spontaneous strain component is also reported for the optimized β and γ phases. Dotted lines are a guide to the eye.

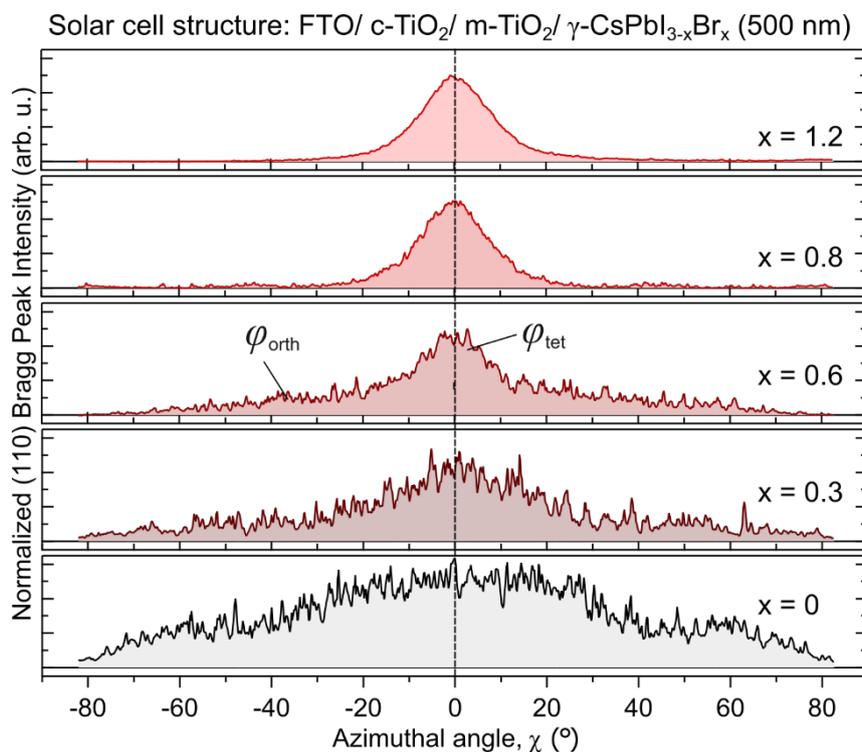

**Fig. S19.** Comparison of the $CsPbI_{3-x}Br_x$ (110) Bragg intensity distribution in the azimuthal domain for varying Br content. The thin films were fabricated atop device-ready substrates (common to halide perovskite solar cells) implementing a mesoporous titanium dioxide (m-$TiO_2$) carrier transport layer with a relatively rough surface (Figure S10).

| Cubic | | | |
|---|---|---|---|
| Plane supercell | Plane primitive α unit cell | Ground state energy p.f.u. (meV) | Relative length free lattice vector (%) |
| (001) | (001) | 7 | 98.7 |
| (101) | (101) | 4 | 98.1 |
| Tetragonal | | | |
| Plane supercell | Plane primitive β unit cell | Ground state energy p.f.u. (meV) | Relative length free lattice vector (%) |
| (001) | (001) | 49 | 99.3 |
| (100) | (110) | 22 | 96.3 |
| (1$\bar{1}$0) | (100) | 11 | 94.9 |
| (101) | (111) | 25 | 97.1 |
| Orthorhombic | | | |
| Plane supercell | Plane primitive γ unit cell | Ground state energy p.f.u. (meV) | Relative length free lattice vector (%) |
| (001) | (001) | 30 | 96.7 |
| (100) | (110) | 24 | 96.4 |
| (1$\bar{1}$0) | (100) | 5 | 93.8 |
| (110) | (010) | 32 | 96.1 |
| (101) | (112) | 23 | 95.8 |

**Table S1.** The difference in ground state energies per formula unit (p.f.u.) between the optimized biaxially strained and the optimized unstrained α/β/γ supercell and the length of the free lattice vector relative to the length of the strained lattice vector. The strained planes are reported both for the supercell and the primitive cell.

Supporting Information References:

1. Eperon, G. E. *et al.* Formamidinium lead trihalide: a broadly tunable perovskite for efficient planar heterojunction solar cells. *Energy & Environmental Science* **7**, 982 (2014).
2. Wang, Q. *et al.* Managing Phase Purities and Crystal Orientation for High-Performance and Photostable Cesium Lead Halide Perovskite Solar Cells. *Sol. RRL* **4**, 2000213 (2020).
3. Dyadkin, V., Pattison, P., Dmitriev, V. & Chernyshov, D. A new multipurpose diffractometer PILATUS@SNBL. *Journal of Synchrotron Radiation* **23**, 825–829 (2016).
4. Ashiotis, G. *et al.* The fast azimuthal integration Python library: *pyFAI*. *Journal of Applied Crystallography* **48**, 510–519 (2015).
5. Rodríguez-Carvajal, J. Recent advances in magnetic structure determination by neutron powder diffraction. *Physica B: Condensed Matter* **192**, 55–69 (1993).
6. Kresse, G. & Furthmüller, J. Efficiency of ab-initio total energy calculations for metals and semiconductors using a plane-wave basis set. *Computational Materials Science* **6**, 15–50 (1996).
7. Kresse, G. & Furthmüller, J. Efficient iterative schemes for ab initio total-energy calculations using a plane-wave basis set. *Physical Review B - Condensed Matter and Materials Physics* **54**, 11169–11186 (1996).
8. Sun, J., Ruzsinszky, A. & Perdew, J. P. Strongly Constrained and Appropriately Normed Semilocal Density Functional. *Phys. Rev. Lett.* **115**, 036402 (2015).
9. Craciun, V., Craciun, D., Wang, X., Anderson, T. J. & Singh, R. K. Transparent and Conducting Indium Tin Oxide Thin Films Grown by Pulsed Laser Deposition at Low Temperatures. *Journal of Optoelectronics and Advanced Materials* **5**, 401–408 (2003).
10. Keem, J. E. & Honig, J. M. *Selected Electrical and Thermal Properties of Undoped Nickel Oxide*. (West Lafayette, IN, USA, 1978).